\documentclass[aps,prd,groupedaddress,eqsecnum]{revtex4}

\usepackage{graphicx}

\begin{document}


\title{A Twisted Kink Crystal in the Chiral Gross-Neveu model}

\author{G\"ok\c ce Ba\c sar and Gerald V.~Dunne}

\affiliation{Physics Department, University of Connecticut, Storrs CT 06269}


\begin{abstract}
We present the detailed properties of a self-consistent crystalline chiral condensate in the massless chiral Gross-Neveu model. We show that a suitable ansatz  for the Gorkov resolvent reduces the functional  gap equation, for the inhomogeneous condensate, to a nonlinear Schr\"odinger equation, which is exactly soluble. The general crystalline solution includes  as special cases all previously known real and complex condensate solutions to the gap equation. Furthermore, the associated Bogoliubov-de Gennes equation is also soluble with this inhomogeneous chiral condensate, and the exact spectral properties are derived. We find an all-orders expansion of the Ginzburg-Landau effective Lagrangian and show how the gap equation is solved order-by-order.

\end{abstract}


\maketitle
\section{\label{intro}Introduction}

In a recent Letter \cite{bd1}, the authors found a new self-consistent {\it crystalline} condensate solution to the gap equation of the massless chiral Gross-Neveu model \cite{gross}. For this complex chiral condensate, the amplitude is periodic and the phase winds by a certain angle over each period. Our approach is based on the observation that a carefully motivated ansatz for the associated Gorkov resolvent reduces the gap equation to a simple ordinary differential equation, an explicitly soluble form of the nonlinear Schr\"odinger equation (NLSE). In general, the gap equation for an {\it inhomogeneous} condensate is a highly nontrivial functional differential equation, so the reduction to the NLSE represents a significant simplification. This resolvent approach is complementary to the inverse scattering approach \cite{dhn,shei}, which also dramatically simplifies the gap equation, but which was not developed for periodic inhomogeneities. Our resolvent method is  based on an extension, to complex and periodic condensates, of the work of Feinberg and Zee \cite{fz,feinberg}. The general solution to the nonlinear Schr\"odinger equation contains all previously known self-consistent condensates of the massless Gross-Neveu models [both chiral and non-chiral] as special cases: the single real kink \cite{dhn}, the single complex kink \cite{shei}, the real kink crystal \cite{thies-gn}, the complex chiral spiral \cite{schon}, and also yields a new complex kink crystal \cite{bd1}.
In the language of condensed matter physics, this crystalline condensate is a new solution of the Eilenberger equation (for the Gorkov resolvent) \cite{eilenberger}, and we also present here the complete exact solution  of the associated Bogoliubov-de Gennes \cite{degennes} equation $H\psi=E\psi$ for this system. The Eilenberger and Bogoliubov-de Gennes equations are fundamental elements of the treatment of a wide class of interacting fermion systems, which are important in many branches of physics, ranging from  particle physics, to  solid state and atomic physics \cite{campbell,rajagopal,casalbuoni,pitaevskii}. Important paradigms include the Peierls-Frohlich model of conduction \cite{peierls}, the Gorkov-Bogoliubov-de Gennes approach to superconductivity \cite{degennes}, and the Nambu-Jona Lasinio (NJL) model of symmetry breaking in particle physics \cite{nambu}.  Here we study a  $1+1$-dimensional version of the NJL model, the ${\rm NJL}_2$ model [also known as the {\it chiral} Gross-Neveu model, $\chi{\rm GN}_2$]. This model has been widely studied as it exhibits asymptotic freedom, dynamical mass generation, and chiral symmetry breaking \cite{gross,dhn, shei,fz}. 

Our primary physical motivation for studying the gap equation of the massless chiral Gross-Neveu model,  the ${\rm NJL}_2$ model, is to understand the $(T, \mu)$ phase diagram of this system. Somewhat surprisingly, the phase diagram of this system is not yet  fully understood. A gap equation analysis based on a homogeneous condensate suggests its phase diagram is the same as its discrete-chiral cousin, the original Gross-Neveu (${\rm GN}_2$) model \cite{gross}, while more recent work finds an inhomogeneous Larkin-Ovchinikov-Fulde-Ferrell (LOFF \cite{loff}) helical complex condensate (``chiral spiral'') below a critical temperature \cite{schon}. In \cite{bd1}, a Ginzburg-Landau approach was used to show that in a region of the phase diagram the free energy is lower for a complex kink crystal, compared to a uniform condensate or a chiral spiral condensate. In this paper we present  the details of the complex crystalline condensate of the ${\rm NJL}_2$ system, and also the {\it exact} spectral properties of fermions in the presence of such a crystalline condensate. This  information will subsequently be used to study the free energy exactly, without resorting to the Ginzburg-Landau approximation.

This state of affairs should be compared and contrasted with the case of the original Gross-Neveu model \cite{gross}, to which we refer as the ${\rm GN}_2$ model, which has a {\it discrete} chiral symmetry rather than the continuous chiral symmetry of the ${\rm NJL}_2$ model. In the ${\rm GN}_2$ model,  the phase diagram has only relatively recently been solved in the particle physics literature, analytically and exactly by a Hartree-Fock analysis \cite{thies-gn}, and numerically on the lattice \cite{deforcrand}. There is a crystalline phase at low temperature and high density, and this phase is characterized by  a periodically inhomogeneous (real) condensate that solves exactly the gap equation.  This phase is not seen in the old phase diagram which was based on a uniform condensate \cite{wolff,treml}. Interestingly, some hints of a problem with the homogeneous condensate assumption were found already in an early lattice study \cite{karsch}. This discrete-chiral  ${\rm GN}_2$ model (with vanishing bare fermion mass) turns out to be mathematically equivalent to several models in condensed matter physics \cite{thies-gn}: the real periodic condensate may be identified with a polaron crystal in conducting polymers \cite{horovitz,braz,campbell}, with a periodic pair potential in quasi 1D superconductors \cite{kuper,mertsching,buzdin}, and with the real order parameter for superconductors in a ferromagnetic field \cite{machida}. The Gross-Neveu models also serve as paradigms of the phenomenon of fermion number fractionalization \cite{jackiw,gw,niemi,heeger}.

Here, we consider the massless chiral Gross-Neveu, or  ${\rm NJL}_2$, model in $1+1$ dimensions  with Lagrangian \cite{gross,shei,dhn}
\begin{equation}
{\mathcal L}=\bar{\psi}\,i\, \partial \hskip -6pt / \,\psi +\frac{g^2}{2}\left[\left(\bar{\psi}\psi\right)^2+\left(\bar{\psi}i\gamma^5 \psi\right)^2\right]\quad .
\label{lag}
\end{equation}
This system has a continuous chiral symmetry under $\psi\to e^{i\gamma^5 \alpha} \psi$. We have suppressed summation over $N$ flavors, which makes the semiclassical gap equation analysis exact in the $N\to\infty$ limit, a limit in which we can consistently discuss chiral symmetry breaking in 2D \cite{witten,affleck}.
The original Gross-Neveu model, the ${\rm GN}_2$ model \cite{gross}, without the pseudoscalar interaction term $\left(\bar{\psi}i\gamma^5 \psi\right)^2$, has a discrete chiral symmetry $\psi\to \gamma^5\psi$.

There are two equivalent ways to find self-consistent static condensates. First, introduce bosonic scalar and pseudoscalar condensate fields, $\sigma$ and $\pi$, which we combine into a complex 
condensate field, defined either through its real and imaginary parts, or via its amplitude and phase:
\begin{eqnarray}
\Delta\equiv  \sigma-i \, \pi \equiv  M\, e^{i\chi} \quad .
\label{complex-cond}
\end{eqnarray}
Integrating out the fermion fields we obtain an effective action for the condensate $\Delta$ as
\begin{equation}
S_{\rm eff}=-\frac{1}{2 g^2}\int |\Delta|^2 -i\,N\,\ln\det\left[i\, \partial \hskip -6pt / \,
-\frac{1}{2}\left(1-\gamma^5\right)\Delta
-\frac{1}{2}\left(1+\gamma^5\right)\Delta^*\right]
\label{effective}
\end{equation}
The corresponding (complex) gap equation is
\begin{equation}
\hskip -2pt \Delta(x)= -2iN g^2 \frac{\delta}{\delta \Delta(x)^*}\ln\det\left[i\, \partial \hskip -6pt / \,
-\frac{1}{2}\left(1-\gamma^5\right)\Delta(x)
-\frac{1}{2}\left(1+\gamma^5\right)\Delta^*(x)\right]
\label{gap}
\end{equation}
If the condensate is constant, as is usually assumed, it is straightforward to evaluate the determinant and solve the gap equation \cite{wolff,treml,casalbuoni}. When the condensate is inhomogeneous this is a much more difficult problem. Dashen, Hasslacher and Neveu \cite{dhn} used inverse scattering  to find kink-like static but spatially inhomogeneous condensates for the gap equation of the ${\rm GN}_2$ model (where there is no pseudoscalar condensate, so $\Delta$ is real). Shei \cite{shei} extended this inverse scattering analysis to the chiral Gross-Neveu model, the ${\rm NJL}_2$ model, and found a spatially inhomogeneous complex kink. A new approach to the inhomogeneous gap equation, based on the resolvent, was developed by Feinberg and Zee \cite{fz} and applied to the kink solutions of both the ${\rm GN}_2$ and ${\rm NJL}_2$ models. For the ${\rm GN}_2$ model, Thies used a Hartree-Fock approach to find a periodic extension of the real kink solution, motivated by analogous inhomogeneous condensates in condensed matter systems \cite{thies-gn}. In  \cite{bd1}, the present authors showed that the complex gap equation (\ref{gap}) can be reduced in an elementary manner to a soluble form of the nonlinear Schr\"odinger equation. The general solution contains all previously known inhomogeneous condensates (real and complex), and yields a new crystalline extension of Shei's complex kink.

A second approach to finding a self-consistent condensate is to solve the relativistic Hartree-Fock problem $H\psi=E\psi$, with single-particle Hamiltonian
\begin{equation}
H=-i\gamma^5 \frac{d}{dx}+\gamma^0 \left[\frac{1}{2}\left(1-\gamma^5\right)\Delta(x)+
\frac{1}{2}\left(1+\gamma^5\right)\Delta^*(x)\right]
\label{ham}
\end{equation}
and subject to the consistency condition
\begin{eqnarray}
\langle\bar{\psi}\psi\rangle-i \langle\bar{\psi} i \gamma^5\psi\rangle=-\Delta/g^2
\label{hf-condition}
\end{eqnarray}
We choose Dirac matrices  $\gamma^0=\sigma_1$, $\gamma^1=-i\sigma_2$, $\gamma^5=\sigma_3$, to emphasize the natural complex condensate combination in (\ref{complex-cond}). Then the single-particle Hamiltonian is
\begin{equation}
H=
\begin{pmatrix}
{-i\frac{d}{dx}&\Delta(x)\cr \Delta^*(x) & i\frac{d}{dx}}
\end{pmatrix}\quad .
\label{ham-bdg}
\end{equation}
This Hamiltonian is also known as the Bogoliubov-de Gennes (BdG)  Hamiltonian, and we will refer to the associated spectral equation
\begin{equation}
H\psi=E\psi
\label{dbdg}
\end{equation}
as the Bogoliubov-de Gennes (BdG) equation.

In Section \ref{functional} we review the reduction of the functional gap equation to the nonlinear 
Schr\"odinger equation (NLSE), and in Sections \ref{real} and \ref{complex} we present the real and complex condensates obtained from solving the NLSE. In Section \ref{solutions} we show that the associated Bogoliubov-de Gennes equation can also be solved exactly, and we derive the exact single particle spectrum and density of states. In Section \ref{hf} we verify the consistency of our solutions by solving the gap equation in the Hartree-Fock approach. An all-orders Ginzburg-Landau expansion of the free energy is presented in Section \ref{allorders}, and we show that the inhomogeneous gap equation is satisfied order-by-order in an interesting and nontrivial way.
In a concluding section we review our results and discuss implications for the phase diagram of the chiral Gross-Neveu model.

\section{Reduction of Functional Gap Equation to Nonlinear Schr\"odinger Equation}
\label{functional}

In this Section we review the reduction \cite{bd1} of the functional gap equation (\ref{gap}) to the nonlinear Schr\"odinger equation. The key quantity in our approach is the coincident limit of Gor'kov Green's function, or the the ``diagonal resolvent'':
\begin{equation}
R(x; E)\equiv \langle x| \frac{1}{H-E} |x\rangle \quad .
\label{res}
\end{equation}
The resolvent (\ref{res}) is clearly a $2\times 2$ matrix.
For a static (but possibly spatially inhomogeneous) condensate, all spectral information is encoded in the resolvent. Indeed, the spectral function characterizing the single-particle spectrum of fermions in the presence of the condensate $\Delta(x)$ is
\begin{equation}
\rho(E)=\frac{1}{\pi}{\rm Im}\,{\rm Tr}_{D,x}\left[R(x;E+i\epsilon)\right]\quad ,
\label{spectral}
\end{equation}
where the trace is a Dirac trace as well as a spatial trace. 

Our first, very simple, observation is that the form of the BdG equation (\ref{dbdg}) places very strong constraints on the possible form of $R(x; E)$. For any static condensate $\Delta(x)$, $R(x; E)$ must satisfy the following algebraic conditions [these are explained in more detail in Appendix A]:
\begin{eqnarray}
R&=&R^\dagger
\label{hermitean}\\
{\rm tr}_D \left(R(x; E)\sigma_3\right) &=&0
\label{condition-1}\\
{\rm det}\, R(x; E)&=&-\frac{1}{4} \quad .
\label{condition-2}
\end{eqnarray}
Furthermore, $R(x; E)$ must satisfy the first-order differential equation 
\begin{eqnarray}
\frac{\partial}{\partial x}R(x; E)\, \sigma_3
&=&i\, \left[
\begin{pmatrix}
{E&-\Delta(x) \cr
\Delta^*(x) & -E}
\end{pmatrix}, R(x; E)\,\sigma_3
\right]
\label{dikii}
\end{eqnarray}
In superconductivity, (\ref{dikii}) is known as the Eilenberger equation \cite{eilenberger,stone}, and in mathematical physics as the Dik'ii equation \cite{dickey}. These conditions (\ref{hermitean})--(\ref{condition-2}), and the Eilenberger equation (\ref{dikii}), all follow from the simple fact \cite{dickey,waxman,stone} that for the one-dimensional BdG equation, which involves derivatives with respect to the single variable $x$, the Green's function can be expressed as a product of two independent solutions to (\ref{dbdg}):
\begin{eqnarray}
R(x; E)=\frac{1}{2iW}\left(\psi_1\psi_2^T+\psi_2\psi_1^T\right)\sigma_1
\label{product}
\end{eqnarray}
where $W$ is the Wronskian of  two independent solutions $\psi_{1,2}$: $
W=i\psi_1^T\sigma_2\psi_2$. 

The next step is to note that the gap equation provides further information about the possible form of the resolvent, and this is enough to motivate a specific ansatz form  \cite{bd1}. There are two ways of viewing the gap equation (\ref{gap}) in terms of the resolvent. First, for a static condensate we can write the log det term in the effective action (\ref{effective}) as minus the grand canonical potential, in terms of the single-particle spectral function $\rho(E)$:
\begin{eqnarray}
-\frac{1}{\beta} \int_{-\infty}^\infty  dE\,\rho(E) \ln\left(1+e^{-\beta(E-\mu)}\right)\quad .
\label{logform}
\end{eqnarray}
All dependence on $\Delta(x)$ resides in the spectral function $\rho(E)$, via (\ref{spectral}). Therefore, inserting this into the gap equation (\ref{gap}), this relates $\Delta(x)$ to the
{\it diagonal} entries of $R(x; E)$. Further,  as a consequence of the condition (\ref{condition-1}), these diagonal entries are equal. So, the simplest natural solution to the gap equation is for the diagonal entries of $R(x; E)$ to be linear in $|\Delta(x)|^2$. A second way to view the gap equation is to evaluate the functional derivative in (\ref{gap}), which for a static condensate leads to:
\begin{eqnarray}
\Delta(x)=-iNg^2{\rm tr}_{D,E}\left[\gamma^0\left({\bf 1}+\gamma^5\right)R(x; E)\right]
\label{gap-2}
\end{eqnarray}
The Dirac trace then relates the {\it off-diagonal} entries of $R(x; E)$ to $\Delta(x)$. Since $R$ is hermitean, these off-diagonal entries are complex conjugates of one another. 

Summarizing, $R(x; E)$ must be a hermitean $2\times 2$ matrix with equal diagonal entries, such that [after the spatial and energy trace] the variation of the diagonal terms is proportional to $\Delta(x)$, and with off-diagonal terms linear in $\Delta(x)$, after the energy trace. This suggests taking the resolvent to be of the form
\begin{equation}
R(x; E)={\mathcal N}(E)
\begin{pmatrix} {
a(E)+|\Delta(x) |^2& b(E)\Delta(x) \cr
b(E) \Delta^*(x)& a(E)+|\Delta(x) |^2 }
\end{pmatrix}
\label{ansatz-1}
\end{equation}
where $a(E)$, $b(E)$ and ${\mathcal N}(E)$ are functions of $E$, independent of $x$, and are to be determined.
However, this ansatz cannot describe inhomogeneous condensates because the only solution of this form consistent with (\ref{condition-2}) is a condensate with constant magnitude, independent of $x$. Indeed, taking $\Delta$ to be constant (and by a global chiral rotation, real), $\Delta=M$, the solution to (\ref{hermitean})--(\ref{dikii}) is simply
\begin{equation}
R(x; E)=\frac{1}{2\sqrt{M^2-E^2}}
\begin{pmatrix} {
E & M \cr
M & E }
\end{pmatrix}
\label{constant-resolvent}
\end{equation}
as is familiar. This example also illustrates that the hermiticity condition (\ref{hermitean}) must of course be interpreted with the appropriate $i\epsilon$ prescription for the energy.

To find inhomogeneous condensates, we suggested in \cite{bd1}  to extend the ansatz (\ref{ansatz-1}) to include a first derivative term in the off-diagonal entries:
\begin{equation}
R(x; E)={\mathcal N}(E)
\begin{pmatrix} {
a(E)+|\Delta(x) |^2& b(E) \Delta(x)-i \Delta^\prime(x) \cr
b(E) \Delta^*(x)+i \Delta^{\prime\,*}(x) & a(E)+|\Delta(x) |^2 }
\end{pmatrix}
\quad .
\label{ansatz}
\end{equation}
This is the simplest extension of (\ref{ansatz-1}) that is consistent with the various algebraic constraints and with the Eilenberger equation (\ref{dikii}). Indeed, substituting the ansatz (\ref{ansatz}) into the Eilenberger equation (\ref{dikii}), we see that the diagonal entry of this equation is identically satisfied, while the off-diagonal entry implies that $\Delta(x)$ must satisfy the following nonlinear Schr\"odinger equation (NLSE) [and its complex conjugate]:
\begin{equation}
\Delta^{\prime\prime} -2|\Delta |^2\,\Delta +i\left(b-2 E\right)\Delta^\prime -2\left(a-E b\right)\Delta=0
\quad .
\label{nlse}
\end{equation}
Two comments are in order. First, it is not immediately obvious that $R(x; E)$ in (\ref{ansatz}) can satisfy the normalization condition (\ref{condition-2}) for an inhomogeneous condensate, since
\begin{eqnarray}
\det  R(x; E)={\mathcal N}^2\left\{ |\Delta |^4-|\Delta^\prime |^2+(2a-b^2)|\Delta|^2+ib\left(\Delta^\prime\Delta^*-\Delta \Delta^{*\prime}\right)+a^2\right\}
\label{det-check}
\end{eqnarray}
Remarkably, the NLSE (\ref{nlse}) implies that $\left({\rm det}\, R(x; E)\right)$ is constant:
\begin{eqnarray}
\frac{d}{dx} \left(\frac{{\rm det}\, R(x; E)}{{\mathcal N}^2}\right)&=&\left(2|\Delta|^2+2a-b^2\right)\left(|\Delta|^2\right)^\prime-\left(\Delta^{\prime\prime}\Delta^{*\prime}+\Delta^\prime\Delta^{*\prime\prime}\right)+ib\left(\Delta^{\prime\prime}\Delta^*-\Delta^{*\prime\prime}\Delta\right)\nonumber\\
&=&0
\label{det-constant}
\end{eqnarray}
where we have used the fact that the NLSE (\ref{nlse}) implies that $\left(\Delta^{\prime\prime}\Delta^{*\prime}+\Delta^\prime\Delta^{*\prime\prime}\right)=\left(2|\Delta|^2+2a-2E b\right)\left(|\Delta|^2\right)^\prime$, and that $\left(\Delta^{\prime\prime}\Delta^*-\Delta^{*\prime\prime}\Delta\right)=-i (b-2E)\left(|\Delta|^2\right)^\prime$. Since
${\rm det}\, R(x; E)$ is constant, the normalization in (\ref{condition-2}) can be achieved by suitable choice of ${\mathcal N}(E)$. Second, while the ansatz (\ref{ansatz}) automatically satisfies the $x$ dependence of the gap equation in its form coming from (\ref{logform}) [because the trace of $R$ is, by construction, linear in $|\Delta|^2$], it doesn't satisfy the other form of the gap equation (\ref{gap-2}), until the energy trace is performed. This is because of the 
$\Delta^\prime(x)$ terms in the off-diagonal. In Section \ref{hf} we show that this form of  the gap equation is indeed satisfied because the coefficient of the $\Delta^\prime(x)$ term vanishes due to the energy trace. 

Thus, we have reduced the very difficult problem of solving the functional gap equation (\ref{gap}) for a self-consistent condensate $\Delta(x)$ to the much simpler problem of solving the NLSE for $\Delta(x)$. In fact, the NLSE (\ref{nlse}) is explicitly soluble, as is discussed in the following sections, in which we describe first the real solutions [relevant for the ${\rm GN}_2$ model], and then the complex solutions [relevant for the ${\rm NJL}_2$ model].

\section{Real solutions of the NLSE}
\label{real}

In this Section we recall the previously known {\it real} solutions to the gap equation, and show how they fit in with the NLSE (\ref{nlse}) and the resolvent form in (\ref{ansatz}).

\subsection{Homogeneous condensate}
\label{homog}

If the condensate is constant, then by a global chiral rotation it can be  taken to be real:
\begin{equation}
\Delta(x)=m
\label{real-constant}
\end{equation}
This clearly satisfies the NLSE (\ref{nlse}), and we find
\begin{eqnarray}
a(E)&=& 2E^2 -m^2 \nonumber \\
b(E)&=&2 E \nonumber \\
{\mathcal N}(E)&=& \frac{1}{4}\frac{1}{E\sqrt{m^2-E^2}}
\label{real-constant-res}
\end{eqnarray}
The spectrum of the associated BdG equation (\ref{dbdg}) is that of a free fermion with mass $m$, with positive and negative energy continua starting at  $E=\pm m$, the mass scale being set by the amplitude of the condensate.

\subsection{Single real kink condensate}
\label{realkink}

A well known nontrivial solution to the gap equation is the single (real) kink \cite{dhn}:
\begin{equation}
\Delta(x)=m\, \tanh\left( m\,x\right)
\label{real-kink}
\end{equation}
This satisfies the NLSE
\begin{equation}
\Delta^{\prime\prime} -2\,\Delta^3  +2m^2\Delta=0\quad ,
\label{real-kink-nlse}
\end{equation}
and so we deduce the exact diagonal resolvent to be of the form (\ref{ansatz}) with
\begin{eqnarray}
a(E)&=& 2E^2-m^2 \nonumber \\
b(E)&=&2 E \nonumber \\
{\mathcal N}(E)&=& \frac{1}{4} \frac{1}{E \sqrt{m^2-E^2}}
\label{real-kink-res}
\end{eqnarray}
The spectrum of the associated BdG equation (\ref{dbdg}) has positive and negative energy continua starting at  $E=\pm m$, together with a single
bound state located at $E=0$, at the center of the gap. This mid-gap zero mode has many important consequences in a variety of branches of physics \cite{jackiw,niemi}.

\subsection{Real kink crystal condensate}
\label{realcrystal}

A periodic array of these real kinks also provides a solution to the gap equation. This solution describes a polaron crystal in polymer physics \cite{horovitz,braz},  a periodic pair potential in inhomogeneous superconductors \cite{kuper,mertsching,machida}, and the crystalline phase of the Gross-Neveu model \cite{thies-gn}. Define (the peculiar looking scaling will become clear below)
\begin{equation}
\Delta(x)=\sqrt{\nu}\,\frac{2 m}{1+\sqrt{\nu}}\,{\rm sn}\left( \frac{2m}{1+\sqrt{\nu}}\,x; \nu \right)
\label{real-kink-crystal}
\end{equation}
where sn is the Jacobi elliptic function \cite{as,ww,akhiezer,lawden} with real elliptic parameter $0\leq \nu\leq 1$. The sn function has period $2{\bf K}(\nu)$, where ${\bf K}(\nu)\equiv \int_0^{\pi/2}\left(1-\nu \,\sin^2t \right)^{-1/2}dt$.  When $\nu=1$ (\ref{real-kink-crystal}) reduces to the single kink condensate in (\ref{real-kink}). The periodic condensate 
(\ref{real-kink-crystal}) satisfies the NLSE
\begin{equation}
\Delta^{\prime\prime} -2\,\Delta^3  +(1+\nu)\left(\frac{2m}{1+\sqrt{\nu}}\right)^2\Delta=0 \quad .
\label{real-kink-crystal-nlse}
\end{equation}
Thus, we deduce the exact diagonal resolvent to be of the form (\ref{ansatz}) with
\begin{eqnarray}
a(E)&=& 2E^2-2m^2\frac{1+\nu}{(1+\sqrt{\nu})^2}\\
b(E)&=&2 E \\
{\mathcal N}(E)&=& \frac{1}{4}\frac{1}{\sqrt{m^2-E^2}\,\sqrt{E^2-m^2(\frac{1-\sqrt{\nu}}{1+\sqrt{\nu}})^2} }
\label{real-kink-crystal-res}
\end{eqnarray}
\begin{figure}
\includegraphics[scale=0.6]{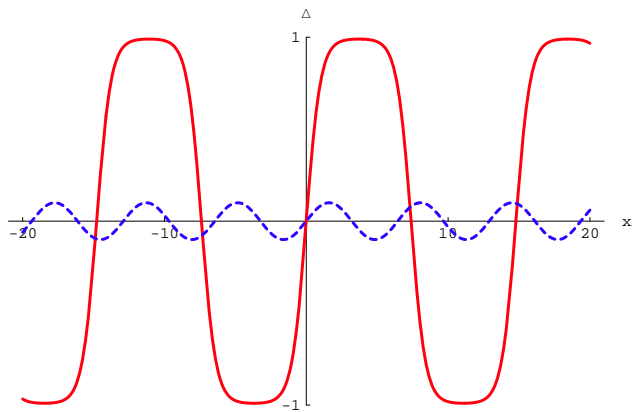}
\caption{The real kink crystal condensate (\ref{real-kink-crystal}) plotted for elliptic parameter $\nu=0.99$ [solid, red curve], and for $\nu=0.1$ [dashed, blue curve]. For small $\nu$ the condensate has the Larkin-Ovchinikov-Fulde-Ferrell (LOFF) form of a small amplitude sinusoidal condensate, while for $\nu\to 1$ the condensate resembles an array of kinks and anti-kinks.}
\label{fig1}
\end{figure}
This periodic condensate is plotted in Figure \ref{fig1}. Note that over the period $x\in [-\frac{{\bf K}(\nu)(1+\sqrt{\nu})}{2m}, \frac{{\bf K}(\nu)(1+\sqrt{\nu})}{2m}]$, the condensate is shaped like a single kink. This reflects the expansion of the Jacobi sn function in terms of an array of periodically displaced tanh functions:
\begin{equation}
{\rm sn}(x; \nu)= \frac{\pi}{2\sqrt{\nu}\, {\bf K}^\prime}\sum_{n=-\infty}^{\infty} (-1)^n \tanh\left(\frac{\pi}{2{\bf K}^\prime}(x- 2n {\bf K})\right)
\label{sn-tanh}
\end{equation}
where we use the standard notation ${\bf K}^\prime(\nu)\equiv {\bf K}(1-\nu)$.
In the infinite period limit ($\nu\to 1$), the interval $[-\frac{{\bf K}(\nu)(1+\sqrt{\nu})}{2m}, \frac{{\bf K}(\nu)(1+\sqrt{\nu})}{2m}]$ maps to the whole real line, and ${\bf K}^\prime\to \pi/2$, and so the kink crystal (\ref{real-kink-crystal}) reduces precisely to the single kink condensate in (\ref{real-kink}).

It is worth noting that this periodic kink crystal (\ref{real-kink-crystal}) can be written in an equivalent, but different looking, form, by use of a Landen transformation \cite{as,ww,akhiezer,lawden} of the Jacobi functions. That is, by rescaling the elliptic parameter $\nu$ together with the argument $m x$, we can write
\begin{equation}
\Delta(x)=m\, \tilde{\nu}\, \frac{{\rm sn}\left( m\,x; \tilde{\nu} \right) {\rm cn}\left( m\,x; \tilde{\nu} \right)}{{\rm dn}\left( m\,x; \tilde{\nu} \right)} \quad ; \quad \tilde{\nu}=\frac{4\sqrt{\nu}}{(1+\sqrt{\nu})^2} \quad .
\label{kink-crystal-landen}
\end{equation}
This is the form in which this periodic kink solution is presented in the work of Thies {\it et al} \cite{thies-gn} on the crystalline phase of the Gross-Neveu model, while the form (\ref{real-kink-crystal}) was used in the condensed matter literature in \cite{kuper,horovitz,braz,mertsching,machida,buzdin}.
\begin{figure}
\includegraphics[scale=0.6]{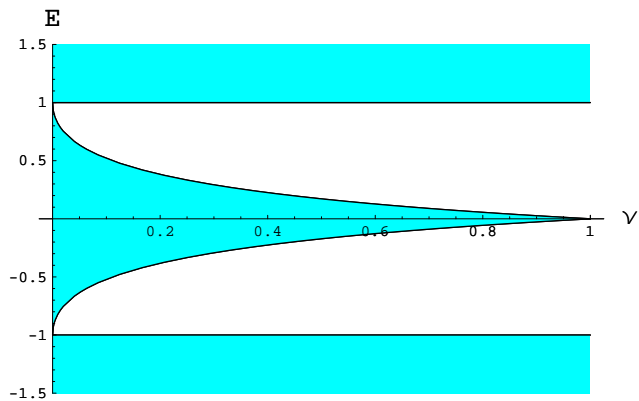}
\caption{The band spectrum of the real kink crystal, showing the positive and negative energy continua and the bound band, as a function of the elliptic parameter $\nu$. The energy is given in units of the scale $m$. The infinite period limit is $\nu\to 1$, where the bound band shrinks to a single bound level at $E=0$, the familiar zero mode of the kink condensate.}
\label{fig2}
\end{figure}
The spectrum of the associated BdG equation (\ref{dbdg}) has positive and negative energy continua starting at  $E=\pm m$, together with a single
bound band in the middle of the gap, with band edges at $E=\pm \left(\frac{1-\sqrt{\nu}}{1+\sqrt{\nu}}\right)m$. This band lies symmetrically in the center of the gap. The spectrum is plotted in Figure \ref{fig2} as a function of the elliptic parameter $\nu$. Notice that there is just one bound band in the energy gap, and when $\nu\to 1$ [the infinite period limit], the bound band at the center of the gap contracts smoothly to the single bound zero mode of the kink condensate.

\section{Complex solutions of the NLSE}
\label{complex}

\subsection{Single plane wave condensate
\label{planewave}}

The simplest complex solution to the NLSE is a single plane wave:
\begin{equation}
\Delta=m \, e^{iqx}
\label{complex-plane}
\end{equation}
This satisfies the NLSE (\ref{nlse}) with
\begin{eqnarray}
a(E)&=& 2\left(E-\frac{q}{2}\right)^2-m^2 \nonumber \\
b(E)&=&2 E -2 q  \quad\qquad \qquad \nonumber \\
{\mathcal N}(E)&=&\frac{1}{4}\frac{1}{(E-q/2) \sqrt{m^2-(E-q/2)^2}}
\label{complex-plane-res}
\end{eqnarray}
This plane wave condensate behaves just like a constant one, but
with the energy shifted by $q/2$, as can be seen by making a local chiral rotation:
\begin{equation}
\Delta\to e^{-iqx} \Delta\qquad , \qquad \psi\to e^{-iqx/2 \gamma_5} \psi \quad .
\label{plane-rotation}
\end{equation}
It is clear from the BdG equation (\ref{dbdg}) that such a transformation has the effect of shifting the entire energy spectrum by $q/2$.
This illustrates an important point:  for complex solutions $\Delta(x)$ of the NLSE, one can always multiply by an arbitrary plane wave phase factor $e^{iqx}$, and this simply corresponds to shifting the entire energy spectrum. 
In general, a  local chiral rotation through angle $\alpha(x)$ leads to a local chemical potential \cite{wilczek,aitchison}, or local electrostatic potential, $A_0(x)=\frac{1}{2}\alpha^\prime(x)$, and therefore a local electric field $E(x)=-\frac{1}{2}\alpha^{\prime\prime}(x)$. For the single plane wave condensate in (\ref{complex-plane}), $\alpha(x)$ is linear in $x$, and so there is no associated electric field.

\subsection{Single complex kink condensate}
\label{complexkink}

Shei \cite{shei} found a solution to the gap equation for the ${\rm NJL}_2$ model, in which both the scalar and  pseudoscalar condensates have a kink-like form:
\begin{eqnarray}
\sigma(x) &=& m\left[\cos^2(\theta/2)+\sin^2(\theta/2) \tanh\left(m\,\sin\left(\theta/2\right)\, x\right)\right] \nonumber \\
\pi(x)&=& -\frac{m}{2}\, \sin(\theta)\left[1-\tanh\left(m\,\sin\left(\theta/2\right)\, x\right)\right]
\label{shei-kink}
\end{eqnarray}
where $\theta\in [0, 2 \pi]$ is a parameter.
\begin{figure}
\includegraphics[scale=0.6]{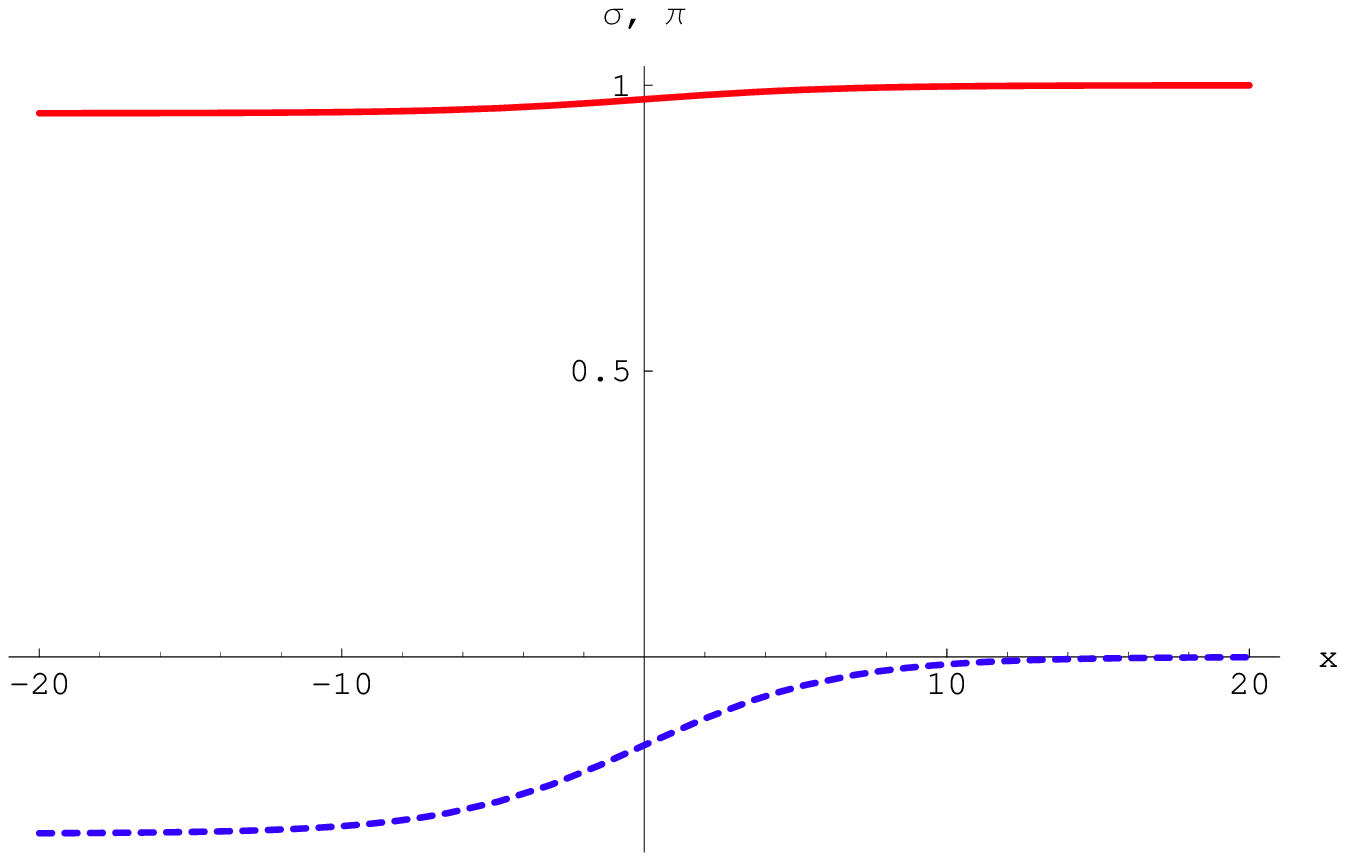}
\includegraphics[scale=0.6]{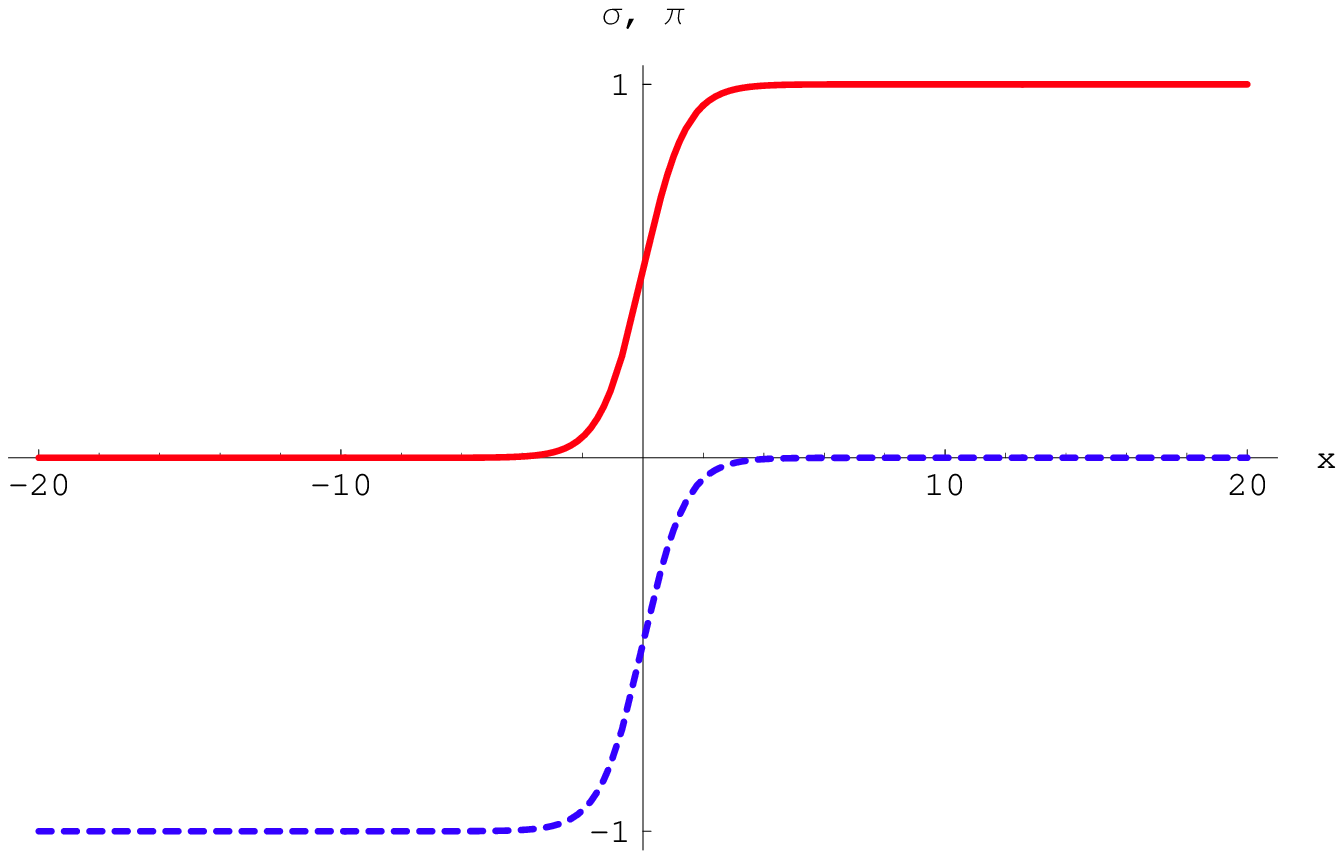}
\includegraphics[scale=0.6]{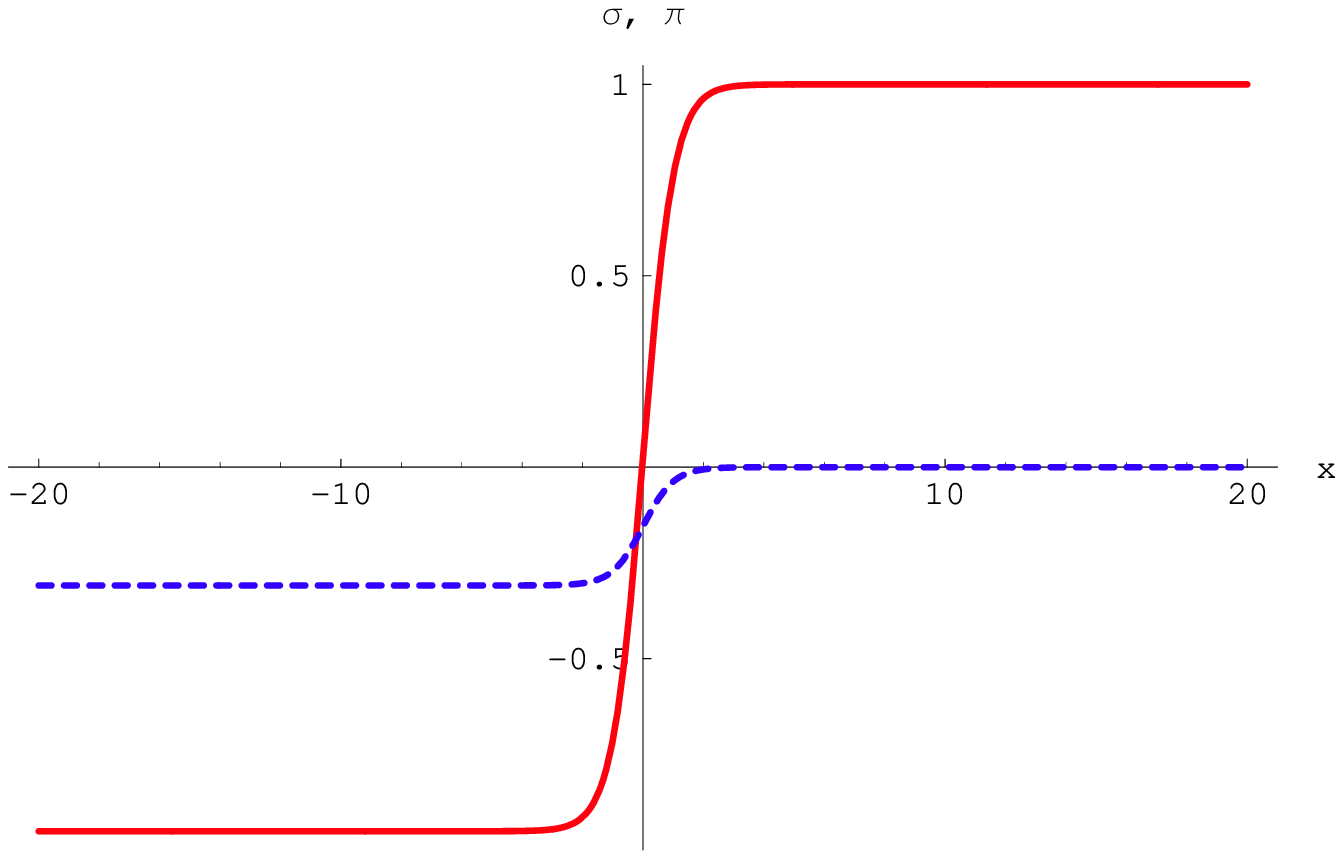}
\caption{Plots of the real and imaginary parts $\sigma(x)$ and $\pi(x)$ of the complex kink  condensate in (\ref{shei-kink}) for three different values of the winding parameter $\theta$. The scalar condensate $\sigma(x)$ [solid, red curves] winds from $m$ at $x=-\infty$, to $m\cos(\theta)$ at $x=+\infty$, while the pseudoscalar kink $\pi(x)$ [dashed, blue, curves] winds from $-m\sin(\theta)$ to 0 as $x$ ranges from  $x=-\infty$ to $x=+\infty$. The plots are for $\theta=\pi/10$, $\theta=\pi/2$ and $\theta=9\pi/10$, and $\sigma$ and $\pi$ are plotted in units of $m$.}
\label{fig3}
\end{figure}
These kinks are plotted in Figure \ref{fig3}. This complex kink (\ref{shei-kink}) has also been extensively studied in the resolvent approach by Feinberg and Zee \cite{fz}. In our analysis it is more natural to combine these into the complex condensate $\Delta(x)=\sigma(x)-i \pi(x)$, [as in (\ref{complex-cond})]:
\begin{equation}
\Delta(x)=m\, \frac{\cosh\left(m\,\sin\left(\theta/2\right)\, x-i\theta/2\right)}{\cosh\left(m\,\sin\left(\theta/2 \right)\, x\right)}\, e^{i\theta/2}\quad .
\label{complex-kink}
\end{equation}
\begin{figure}
\includegraphics[scale=0.6]{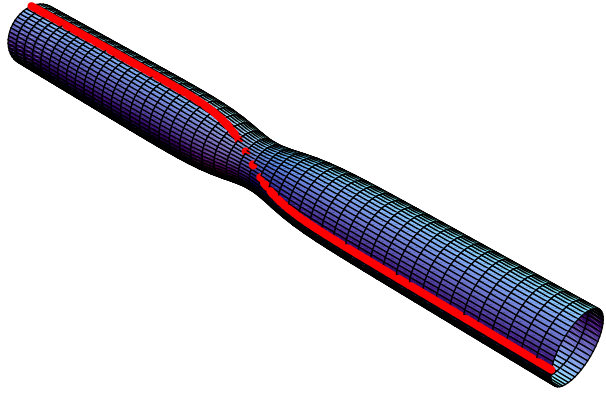}
\caption{Plot of  the complex kink  condensate (\ref{complex-kink}), for $\theta=3\pi/2$, illustrating how the kink winds around zero without the amplitude vanishing. The kink is the solid [red] line, and the surface is shown simply to illustrate that both the amplitude and the phase are changing.}
\label{fig4}
\end{figure}
This complex form is plotted in Figure \ref{fig4}. This illustrates  the role of the parameter $\theta\in [0, 2 \pi]$ as the net rotation angle of the kink as $x$ goes from $-\infty$ to $+\infty$:
\begin{equation}
\Delta(x=+\infty)=e^{-i\theta}\Delta(x=-\infty)
\label{complex-kink-rotation}
\end{equation}
Observe that when $\theta=\pi$, the complex kink (\ref{complex-kink}) is in fact real, and reduces to the familiar real kink solution in (\ref{real-kink}); this real kink changes its sign [i.e., rotates through $\pi$] in passing from $x=-\infty$ to $x=+\infty$.  Another useful representation of this kink is in terms of the magnitude and phase  $\Delta(x)=M(x)e^{i\chi(x)}$:
\begin{eqnarray}
M^2(x) &=&   m^2\left[1-\sin^2\left(\theta/2\right){\rm sech}^2\left(m\,\sin\left(\theta/2\right)\, x\right)\right]
\label{complex-kink-amp}\\
\chi(x)&=& {\rm arctan}\left(\frac{\sin\theta}{\cos\theta+e^{2mx\sin(\theta/2)}}\right)
\label{complex-kink-phase}
\end{eqnarray}
\begin{figure}
\includegraphics[scale=0.6]{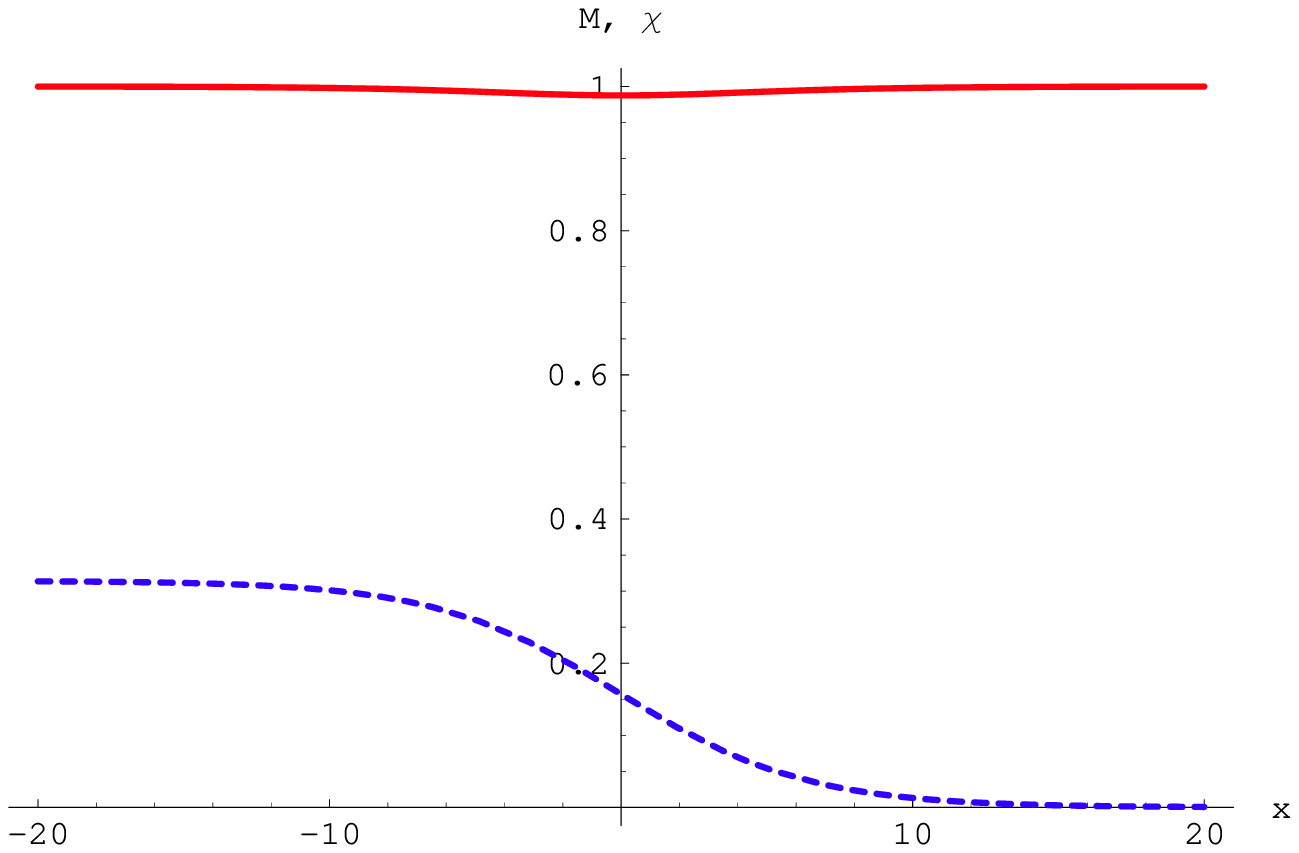}
\includegraphics[scale=0.6]{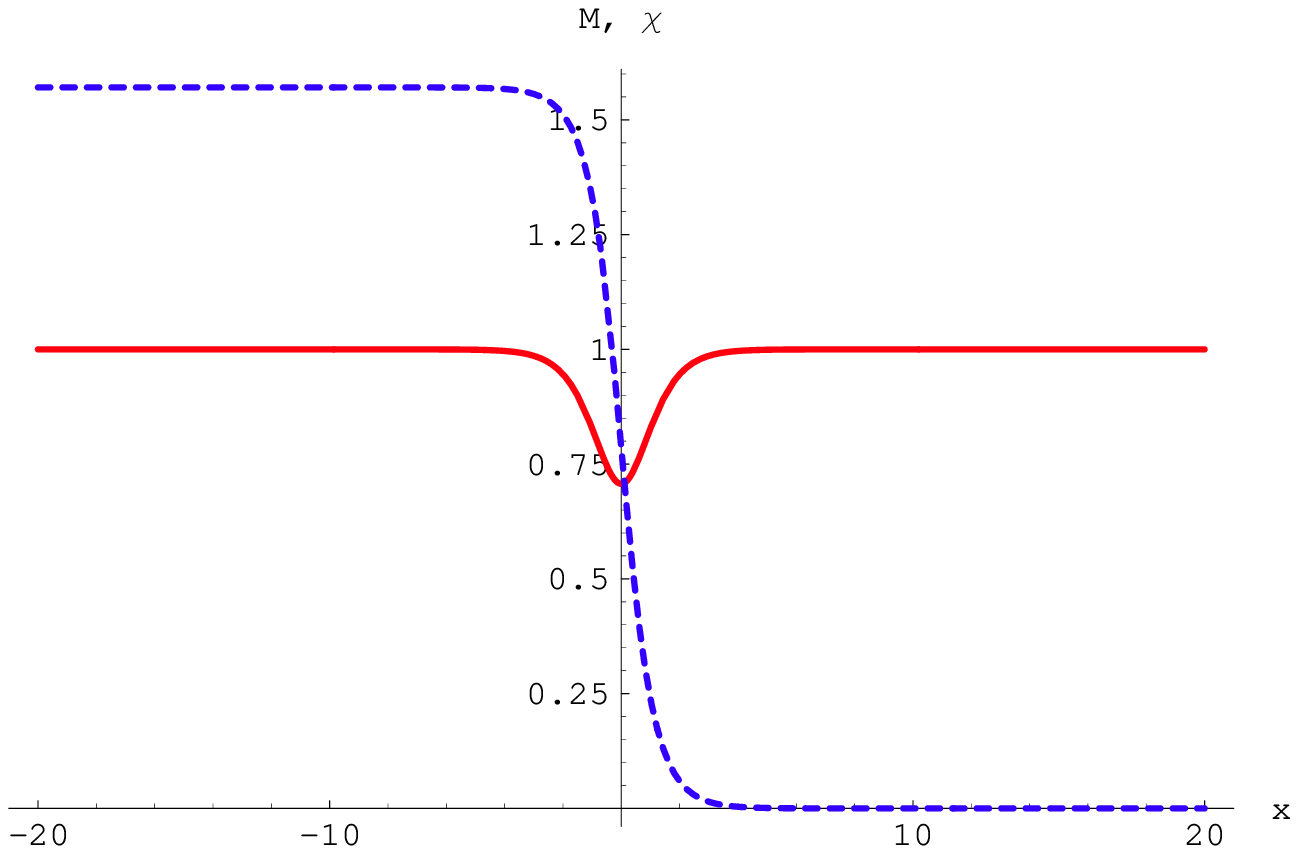}
\includegraphics[scale=0.6]{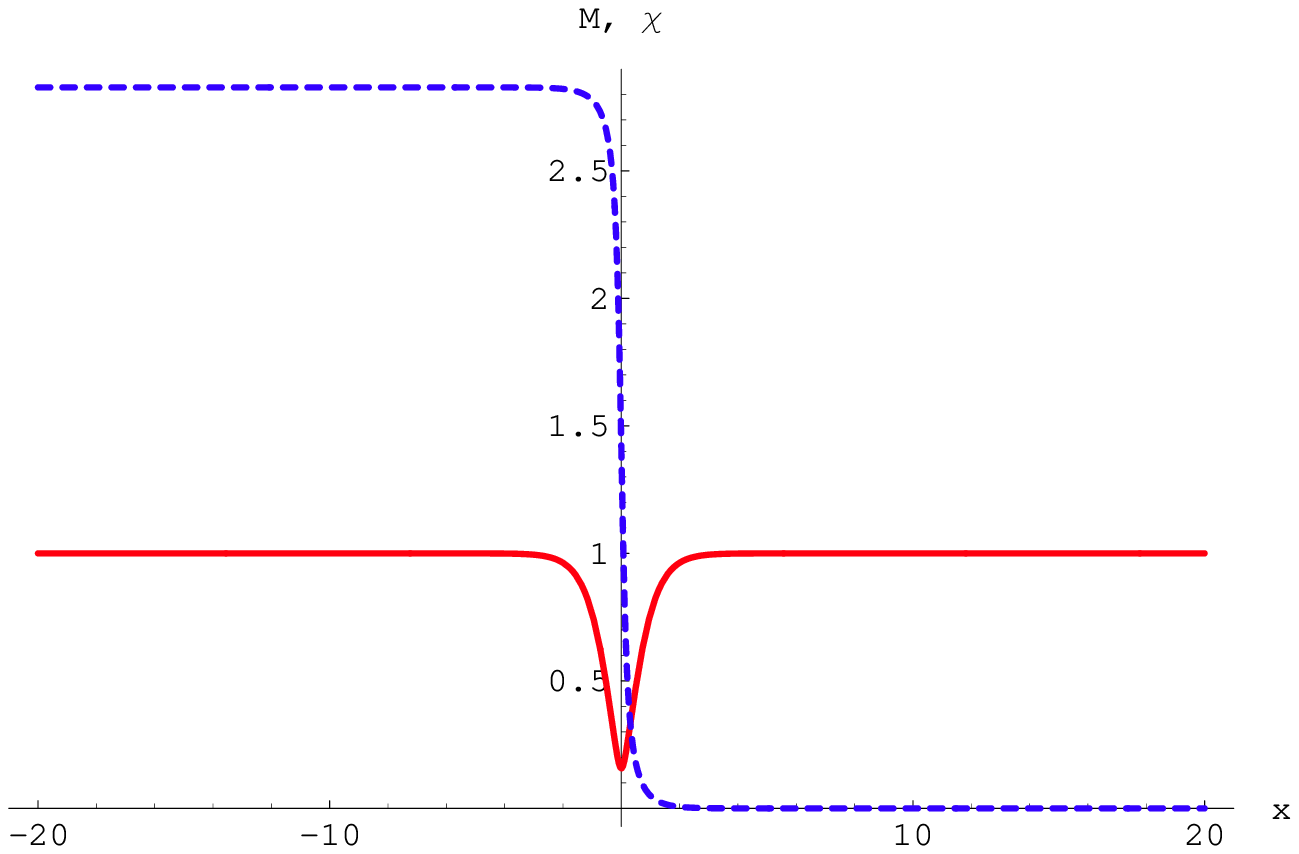}
\caption{Plots of the amplitude $M$ and phase $\chi$ of the 
complex kink  condensate (\ref{complex-kink}), for three different values of the winding parameter $\theta$. The  condensate amplitude $M(x)$ [solid, red curves] approaches $m$ at  $x=\pm\infty$, and equals $m\cos(\theta/2)$ at the kink center $x=0$. The phase $\chi(x)$ [dashed, blue, curves] winds from $\theta$ to 0 as $x$ ranges from  $x=-\infty$ to $x=+\infty$. The plots are for $\theta=\pi/10$, $\theta=\pi/2$ and $\theta=9\pi/10$.}
\label{fig5}
\end{figure}
The complex kink condensate (\ref{complex-kink}) satisfies the NLSE:
\begin{equation}
\Delta^{\prime\prime} -2\,|\Delta|^2\,\Delta -2\,i\, m\, \cos\left(\theta/2\right)\, \Delta^\prime +2m^2\Delta=0
\label{complex-kink-nlse}
\end{equation}
From this NLSE, we deduce the exact diagonal resolvent to be of the form (\ref{ansatz}) with
\begin{eqnarray}
a(E)&=& 2E^2-2\,m\, \cos(\theta/2)E-m^2\nonumber \\
b(E)&=&2 E -m\, \cos(\theta/2) \nonumber \\
{\mathcal N}(E)&=&\frac{1}{4(E-m\,\cos\left(\theta/2\right)) \sqrt{m^2-E^2}}
\label{complex-kink-res}
\end{eqnarray}
\begin{figure}
\includegraphics[scale=0.6]{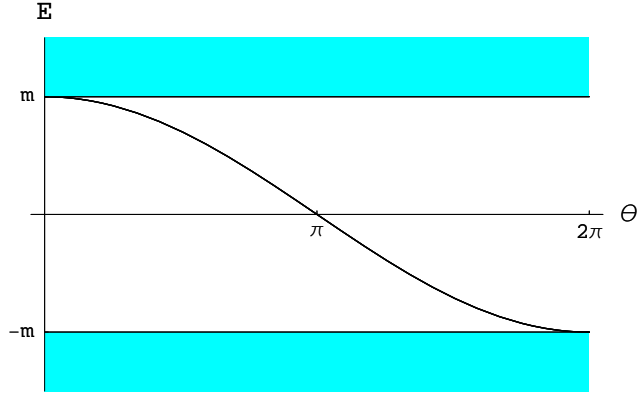}
\caption{Plot of the fermion single-particle spectrum for the single complex kink  (\ref{complex-kink}), as a function of the winding parameter $\theta$. Note that for $\theta=\pi$ [when the condensate is real] the bound state is at $0$, but for all other values of $\theta$ the bound state lies asymmetrically in the gap.}
\label{fig6}
\end{figure}

The spectrum of the associated BdG equation (\ref{dbdg}) has positive and negative energy continua starting at  $E=\pm m$, together with a single
bound state located at $E=m\,\cos\left(\theta/2\right)$. When $\theta=\pi$, where this complex kink reduces to the standard real kink, the bound state is once again a zero mode. But for other values of $\theta$ the single bound state lies asymmetrically inside the gap, as plotted in Figure \ref{fig6} . As $\theta$ goes from $0$ to $2\pi$, one state moves from the positive to the negative energy continuum.  As is clear from the previous subsection, we can always multiply the complex kink solution (\ref{complex-kink}) by a plane-wave factor $e^{iqx}$, which has the net effect of displacing the fermion spectrum by $q/2$, with the corresponding simple modifications to the resolvent functions $a(E)$, $b(E)$ and ${\mathcal N}(E)$ in (\ref{complex-kink-res}).

At this stage we have shown that Shei's complex kink condensate (\ref{complex-kink}) solves the NLSE, and we have found the corresponding exact diagonal resolvent (\ref{ansatz}) with $a(E)$, $b(E)$ and ${\mathcal N}(E)$ given in (\ref{complex-kink-res}). This agrees with the spectral properties derived from inverse scattering \cite{shei}. Shei further showed \cite{shei} that this complex condensate solves the gap equation provided a further restriction is applied to the winding parameter $\theta$. This condition states that $\theta/(2\pi)$ is equal to the filling fraction $\frac{n}{N}$, in the large flavor  limit, of the single bound state in the gap by $n$ flavors, with $\frac{n}{N}$ fixed as $N\to\infty$ \cite{shei,fz}:
\begin{equation}
\frac{\theta}{2\pi}=\frac{n}{N}
\label{shei-condition}
\end{equation}
In Section \ref{hf}  we show that  in our approach this same condition arises from demanding that the coefficient of the $\Delta^\prime(x)$ term in (\ref{gap-2}) vanishes after the energy trace, a necessary requirement to satisfy the gap equation.

\subsection{Complex kink crystal condensate}
\label{complexcrystal}

\begin{figure}
\includegraphics[scale=0.6]{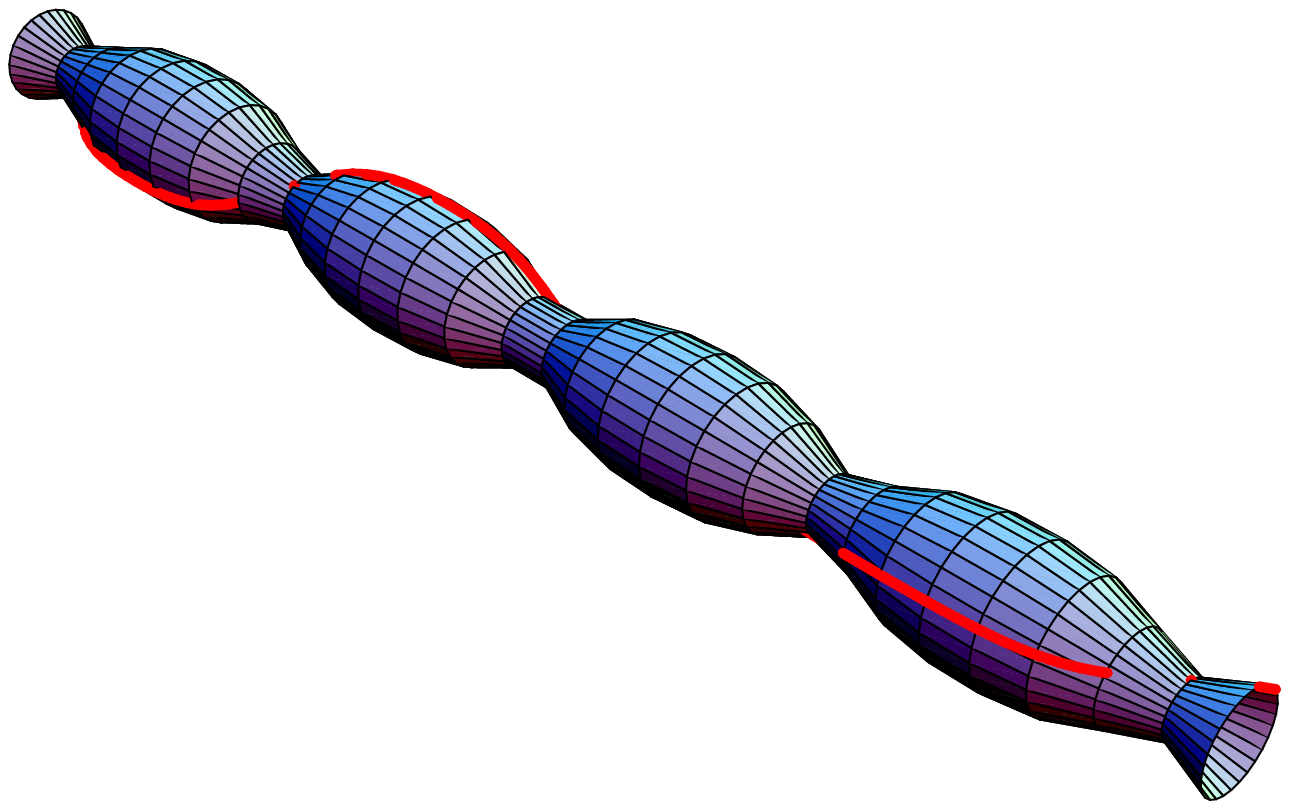}
\caption{Plot of the complex kink  crystal condensate (\ref{complex-kink-crystal}), for $\nu=0.8$ and $\theta=3{\bf K}(0.2)/2$, illustrating how the kink winds around zero each period, without the amplitude vanishing. The kink is the solid [red] line, and the surface is shown simply to illustrate that both the amplitude and the phase are changing over each period.}
\label{fig7}
\end{figure}

A new complex condensate was presented in \cite{bd1}. This new solution is  a periodic array of Shei's complex  kink (\ref{complex-kink}). Physically, it is associated with a crystalline phase of the ${\rm NJL}_2$ system \cite{bd1}, just as the real kink crystal condensate (\ref{real-kink-crystal}) is associated with a crystalline phase of the ${\rm GN}_2$ system \cite{thies-gn}. Up to a plane wave factor [as in Section \ref{planewave}], this complex crystalline condensate is the general solution to the NLSE (\ref{nlse}), and all other solutions [both real and complex] can be obtained from it by suitable choices of parameters. This solution
can be written \footnote{Note that we have made a Jacobi imaginary transformation $x\to i x$, which involves interchanging the real and imaginary periods, relative to the notation in \cite{bd1}, in order to be consistent with the elliptic parameter  conventions of \cite{thies-gn}.} in terms of  Weierstrass elliptic functions [or, alternatively but equivalently, in terms of Jacobi theta functions]
\begin{eqnarray}
\Delta(x)&=&-A \frac{\sigma(A\, x+i{\bf K}^\prime -i\theta/2)}{\sigma(A\, x+i{\bf K}^\prime)\sigma(i\theta/2)}
\,\exp\left[i A\, x \left(-i\,\zeta(i\theta/2)+i\,{\rm ns}(i\theta/2)\right)+i\,\theta \eta_3/2\right] \quad
\label{complex-kink-crystal}
\end{eqnarray}
The parameter $A$ sets the scale of the condensate and its length scale:
\begin{eqnarray}
A=A(m, \theta, \nu)\equiv -2i\,m\, {\rm sc}\left(i\theta/4\right) {\rm nd}\left(i\theta/4\right)
\label{factor}
\end{eqnarray}
where sc=sn/cn and nd=1/dn are Jacobi elliptic functions \cite{as,ww}.
The functions $\sigma$ and $\zeta$ are the Weierstrass sigma and zeta functions \cite{as,ww,akhiezer,lawden}, some relevant properties of which are given in Appendix B. We have chosen real and imaginary half-periods: $\omega_1={\bf K}(\nu)$, and $\omega_3=i\,{\bf K}^\prime\equiv i{\bf K}(1-\nu)$.
Both periods are therefore controlled by the single [real] elliptic parameter $0\leq \nu\leq 1$.
Also, $\eta_3\equiv\zeta(i{\bf K}^\prime)$ is purely imaginary. The parameter $\theta\in [0, 4{\bf K}^\prime(\nu)]$ is related to the angle through which the condensate rotates in one period $L=2{\bf K}/A$ :
\begin{equation}
\Delta(x+L)=e^{2i\varphi} \Delta(x) 
\label{winding-1}
\end{equation}
where the angle $\varphi$ is a function of $\theta$ and $\nu$
\begin{equation}
\varphi= {\bf K}\left(-i\,\zeta(i\theta/2)+i\,{\rm ns}(i\theta/2)-\frac{\eta \theta}{2{\bf K}}\right)
\label{winding-2}
\end{equation}
Here we used the quasi-periodicity property (\ref{sigmaperiod}) of the $\sigma$ function.
Note that $\varphi$ and $\eta\equiv\zeta({\bf K})$ are real, and when $\nu\to 1$, we have $\varphi\to -\theta/2$.
This crystalline complex kink is plotted in Figure \ref{fig7} showing the winding of the kink over a period.

\begin{figure}
\includegraphics[scale=0.6]{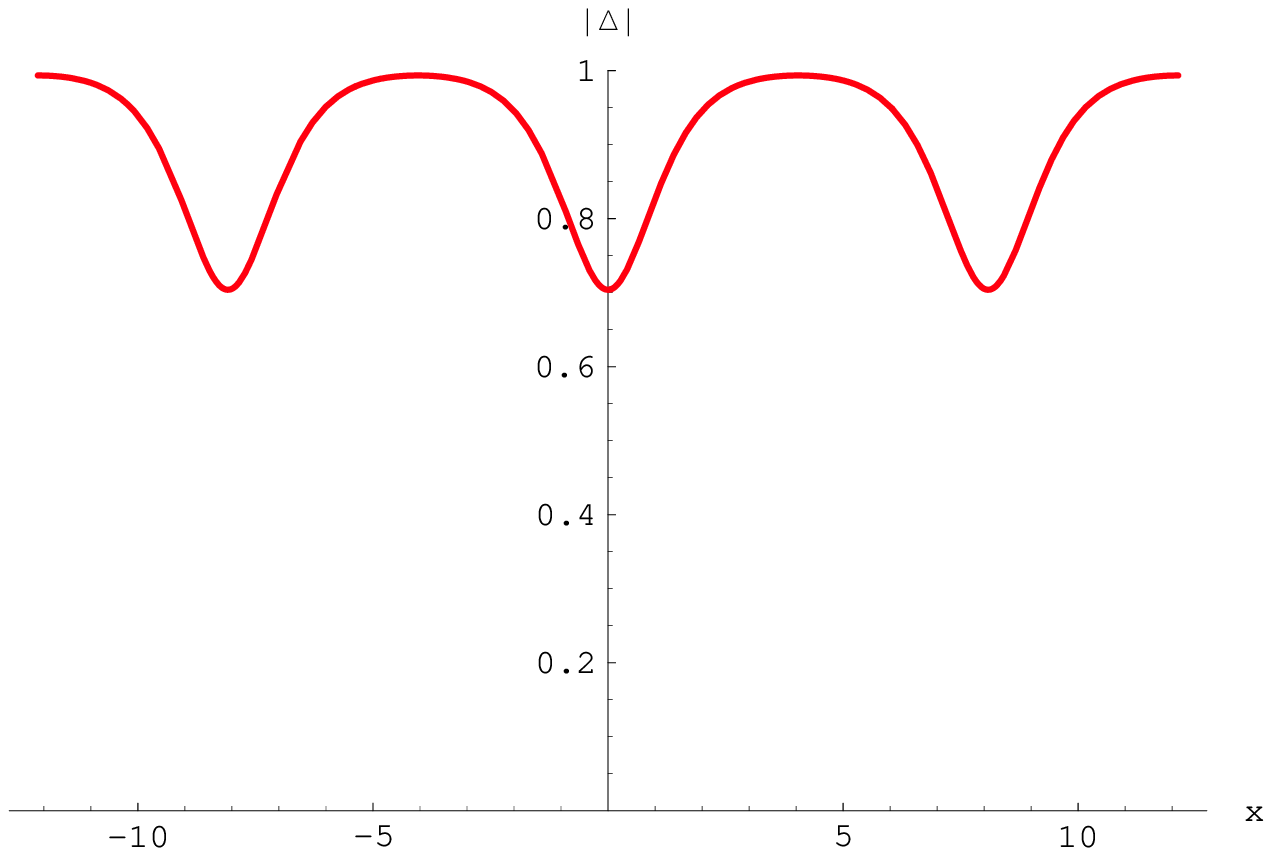}
\includegraphics[scale=0.6]{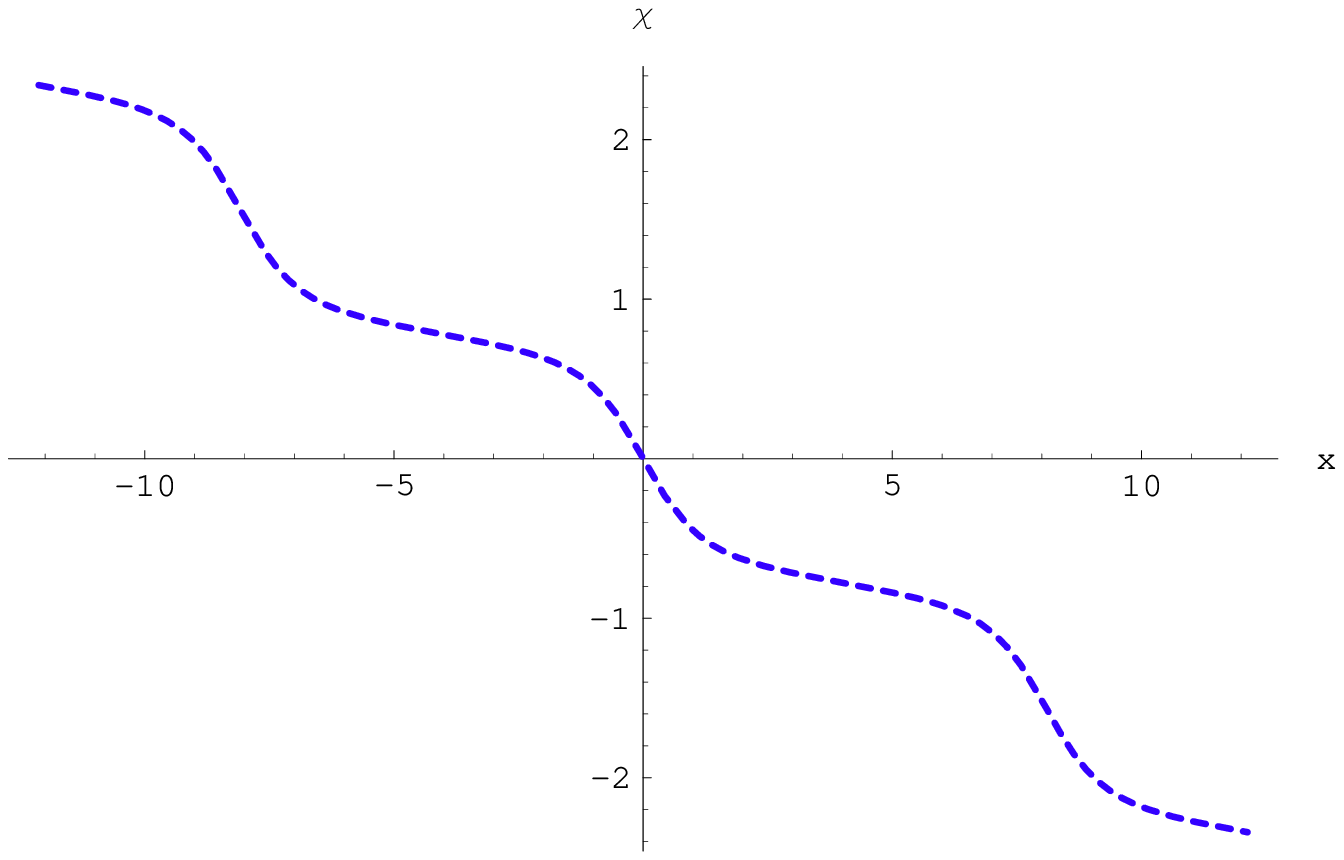}
\caption{Plots of the amplitude $M(x)$ and phase $\chi(x)$ of the 
complex kink  crystal condensate (\ref{complex-kink-crystal}), for $\theta=1.6$ and $\nu=0.95$.}
\label{fig8}
\end{figure}

It is also useful to visualize the condensate (\ref{complex-kink-crystal})  in terms of its amplitude and phase: $\Delta=M\, e^{i\chi}$. The modulus squared is a bounded periodic function, with period $2{\bf K}/A$:
\begin{equation}
M^2\equiv |\Delta (x)|^2 = A^2\left({\mathcal P}\left(A x+i {\bf K}^\prime \right)-{\mathcal P}\left(i\theta/2\right)\right) \quad
\label{complex-kink-crystal-amp}
\end{equation}
Here we used the quasi-periodicity property (\ref{sigmaperiod}) of the $\sigma$ function, together with the product identity (\ref{addition1}) relating the $\sigma$ and ${\mathcal P}$ functions.
The phase $\chi$ can be expressed as
\begin{equation}
\chi(x) = A (-i\,\zeta(i\theta/2)+i\,{\rm ns}(i\theta/2))x+\frac{i}{2}\ln \left(\frac{\sigma(A x+i{\bf K}^\prime +i\theta/2)}{\sigma(A x+i{\bf K}^\prime -i\theta/2)}\right)+\frac{\eta_3 \theta}{2} 
\label{complex-kink-crystal-phase}
\end{equation}
The amplitude and phase are plotted in Figure \ref{fig8}. Note that the amplitude is periodic while the phase changes by $2\varphi$ over each period.

\begin{figure}
\includegraphics[scale=0.6]{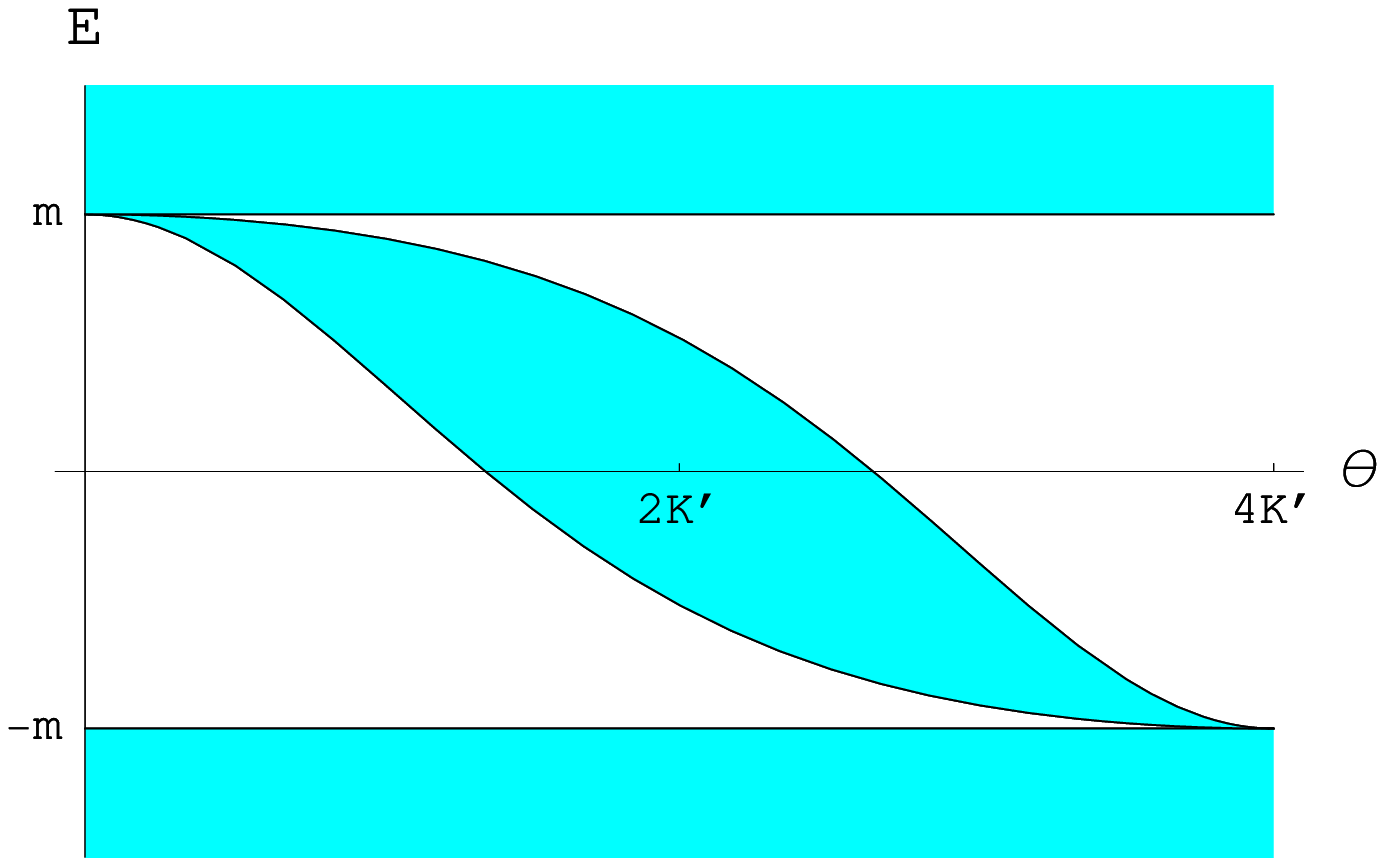}
\includegraphics[scale=0.6]{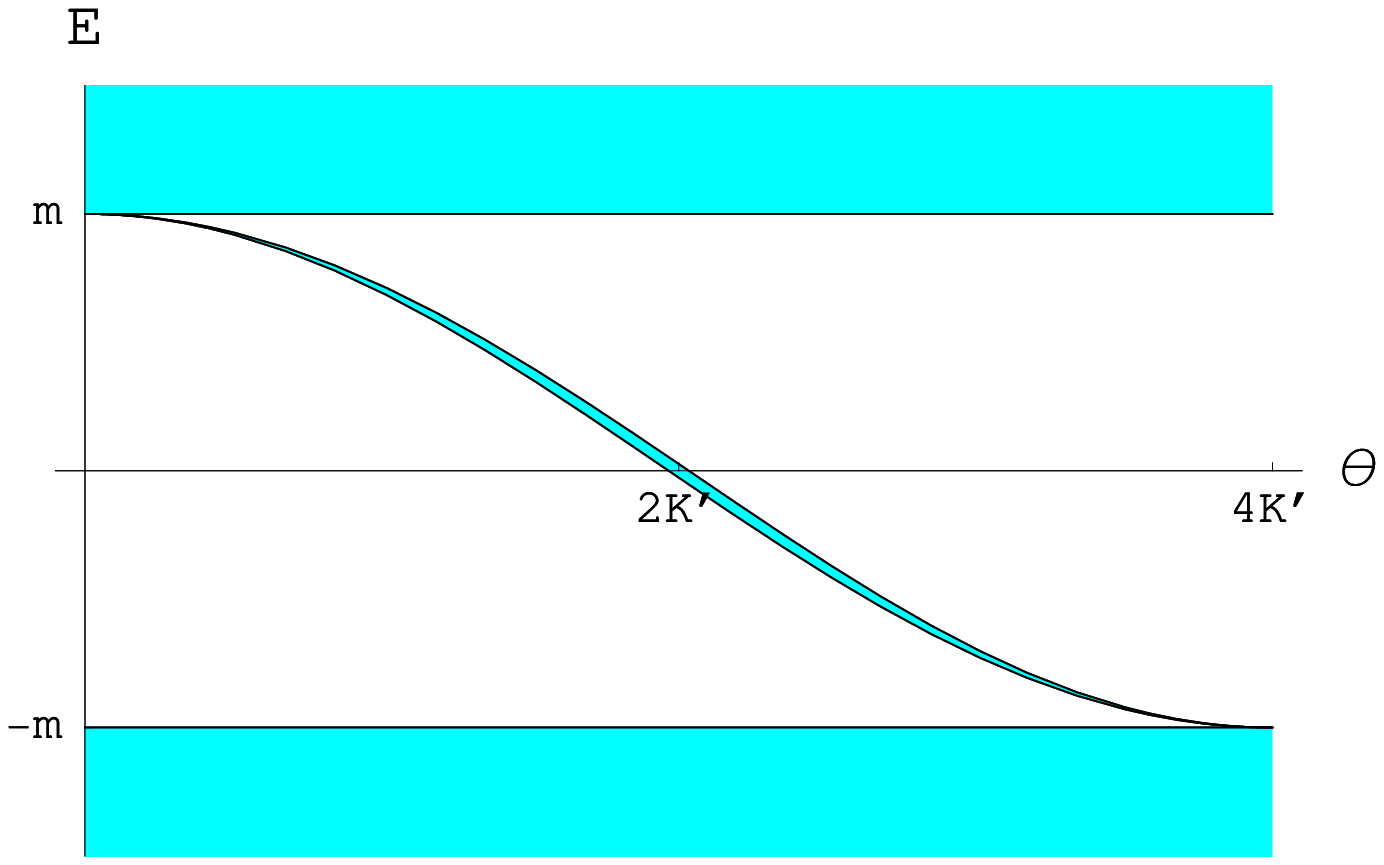}
\caption{Plots of the single-particle fermion spectrum for the complex kink  crystal condensate (\ref{complex-kink-crystal}), for $\nu=0.1$ [first plot] and $\nu=0.9$ [second plot], as a function of the winding parameter $\theta$. Note that for $\theta=2{\bf K}^\prime$ [when the condensate is real] the band is centered symmetrically about $E=0$, but for all other values of $\theta$ the band lies asymmetrically in the gap. In the infinite period limit, $\nu\to 1$, the bound band shrinks to a single bound state, and its $\theta$ dependence reduces to that depicted in Figure \ref{fig6} for the single complex kink condensate.}
\label{fig9}
\end{figure}

The complex crystalline  condensate in (\ref{complex-kink-crystal}) satisfies the NLSE:
\begin{equation}
\Delta^{\prime\prime} -2\,|\Delta|^2\,\Delta -i\,\left(2A\,i\,{\rm ns}(i\theta/2)\right) \Delta^\prime-A^2\left(3\,{\mathcal P}(i\theta/2)-{\rm ns}^2(i\theta/2)\right)\Delta=0
\label{complex-kink-crystal-nlse}
\end{equation}
Comparing this equation with the NLSE (\ref{nlse}) we can extract the functions $a(E)$, $b(E)$ and ${\mathcal N}(E)$ appearing in (\ref{ansatz}), thereby determining the {\it exact} diagonal resolvent.
To express these functions in a compact form, we define some properties of the associated fermionic spectrum for the BdG equation (\ref{dbdg}). This spectrum has positive and negative energy continua starting at  $E=\pm m$, together with a single bound band in the  gap, as depicted in Figure \ref{fig9}. In contrast to the case for the real kink crystal in Section \ref{real}, here the bound band is not centered in the middle of the gap, but is displaced from the center. The parameter $\theta$ characterizes this asymmetry in the spectrum. The band edges are functions of both the winding angle $\theta$ and the elliptic parameter $\nu$:
\begin{eqnarray}
E_1&=&-m \nonumber \\
E_2&=&m(-1+2\, {\rm nc}^2(i\theta/4; \nu)) \nonumber \\
E_3&=&m(-1+2\, {\rm nd}^2(i\theta/4; \nu))\nonumber \\
E_4&=&+m
\label{band-edges}
\end{eqnarray}
In the infinite period limit [$\nu\to 1$], the band contracts to a single bound state, with $E_2=E_3=m \cos\left(\theta/2\right)$, and this is precisely the bound state of the single complex kink, as shown in Figure \ref{fig6}.
At finite period, but when $\theta=2 {\bf K}^\prime(\nu)$, we find $E_2=-E_3=- \left(\frac{1-\sqrt{\nu}}{1+\sqrt{\nu}}\right) m$, and the band is centered symmetrically about $0$; this is precisely the band spectrum of the real kink array in Section \ref{real}.
Thus, we can roughly think of the parameter $\theta$ as setting the offset location of the band inside the gap, while the parameter $\nu$, together with $\theta$, plays the role of determining the period of the crystal, and hence the width of the band.
In terms of the band edges, the resolvent functions $a(E)$, $b(E)$ and ${\mathcal N}(E)$ that appear in the resolvent (\ref{ansatz}) take the following simple form \footnote{In general, when we allow for an overall shift of the energy spectrum by including an extra plane-wave phase in the condensate $\Delta(x)$, these functions can be written as
$a(E)=2E^2-\left(\sum_{j=1}^4 E_j\right)E+\frac{1}{8}\left[\left(\sum_{j=1}^4E_j\right)^2-\sum_{i<j}^4(E_i-E_j)^2\right]$, $b(E)=2E-\left(\sum_{j=1}^4 E_j\right)$, and ${\mathcal N}(E)=1/\left(4\sqrt{\prod_{j=1}^4 (E-E_j)}\right)$.}:
\begin{eqnarray}
a(E)&=&2E^2-(E_2+E_3)E-\frac{(E_2-E_3)^2}{4}-m^2 \nonumber\\
b(E)&=&2 E-(E_2+E_3) \nonumber\\
{\mathcal N}(E)&=&\frac{1}{4\sqrt{(m^2-E^2)(E-E_2)(E-E_3)}}
\label{complex-kink-crystal-res}
\end{eqnarray}
Here the normalization ${\mathcal N}(E)$ is fixed by the property ${\rm det}\,R=-\frac{1}{4}$.

\subsubsection{Solution of NLSE by separation into amplitude and phase}
\label{nlse1}

We now comment on two ways to derive this solution to the NLSE (\ref{nlse}). The first is to separate the NLSE  into two equations, one for the amplitude and one for the phase. Writing $\Delta(x)=M(x)e^{i\chi(x)}$, and considering $(\Delta^*\Delta^{\prime\prime}-\Delta^{\prime\prime*}\Delta)$, we immediately find that the phase is related to the amplitude by:
\begin{equation}
\chi^\prime(x)=-\frac{1}{2}\left(b-2E\right)+\frac{C_1}{M^2(x)} \qquad
\label{general-phase}
\end{equation}
where $C_1$ is a constant. Note that this is indeed true for all the complex condensate solutions discussed above. For the crystalline complex  solution in (\ref{complex-kink-crystal-amp}, \ref{complex-kink-crystal-phase}) the relation (\ref{general-phase})  can be verified using the Weierstrass function properties listed in Appendix B, in particular equation (\ref{addition4}).

Next, considering $(\Delta^*\Delta^{\prime\prime}+\Delta^{\prime\prime*}\Delta)$, we find the following nonlinear equation for $M^2$:
\begin{equation}
\left(\left(M^2\right)^\prime\right)^2=4 M^6+2\left(4(a-E b)-\frac{1}{2}\left(b-2E\right)^2\right) M^4+C_2M^2-4C_1^2
\label{general-amp}
\end{equation}
where $C_2$ is another constant. From the form of this equation, comparing with the equation (\ref{Pdiff}) for the Weierstrass ${\mathcal P}$ function \cite{as,ww,akhiezer,lawden}, we recognize the solution to be of the form
\begin{equation}
M^2=A^2\left[{\mathcal P}(A x+i{\bf K}^\prime) -{\mathcal P}(C_3)\right]
\label{p-sol}
\end{equation}
The shift by $i{\bf K}^\prime$ ensures that $M^2(x)$ is bounded for $x$ on  the real  axis. This explains the result in (\ref{complex-kink-crystal-amp}), identifying $C_3=i \theta/2$.
The constants $C_1$ and $C_2$ are related to $C_3$ as
\begin{eqnarray}
C_1&=& -\frac{i}{2}A^3{\mathcal P}^\prime(i\theta/2)\nonumber\\
C_2&=&2A^4\,{\mathcal P}^{\prime\prime}(i\theta/2)
\label{cs}
\end{eqnarray}

Given this solution for the amplitude, the phase follows from (\ref{general-phase}), using the integration formula \cite{as}:
\begin{equation}
\frac{1}{A}\int \frac{{\mathcal P}^{\prime}(C_3)}{{\mathcal P}(Ax+i{\bf K})-{\mathcal P}(C_3)}dx= \ln \left(\frac{\sigma(Ax+i{\bf K}-C_3)}{\sigma(Ax+i{\bf K}+C_3)}\right)+2(A x+i{\bf K})\zeta(C_3)
\label{integral}
\end{equation}
Notice that the phase can always be shifted by $q\,x$, which amounts to multiplying the solution by a plane wave factor, as discussed in Section \ref{planewave}.
In our solution (\ref{complex-kink-crystal}), the plane wave factor has been chosen so that the associated fermion spectrum has the form shown in Figure \ref{fig9}. An additional plane wave factor in $\Delta(x)$ displaces this entire spectrum by a constant.

\subsubsection{Solution of NLSE as periodic array of complex kinks}
\label{nlse2}

Another way to derive the general periodic complex solution to the NLSE is to make an educated guess for a periodic array of Shei's complex kink solution (\ref{complex-kink}), using known properties of the Weierstrass functions. Shei's complex kink condensate(\ref{complex-kink}) can be written as
\begin{equation}
\Delta(x)=m \frac{\sinh\left(m\sin \left(\theta/2\right) x+i\pi/2-i \theta/2\right)}{\sinh\left(m \sin \left(\theta/2 \right) x+ i\pi/2\right)}e^{i\theta/2}
\label{shei-2}
\end{equation}
This can be made quasi-periodic along the real $x$ axis by generalizing the hyperbolic sine function to its doubly-periodic form, which is the Weierstrass sigma function. Thus, we are led to try the form
\begin{equation}
\Delta(x)=A e^{i\theta\,\eta_3/2}\frac{\sigma\left(A x+i{\bf K}^\prime-i\theta/2\right)}{\sigma\left(A x+i{\bf K}^\prime \right)\,\sigma\left(i\theta/2 \right)}e^{i\lambda x}
\label{guess}
\end{equation}
where we have included a possible plane wave factor $e^{i\lambda\,x}$, to be determined, and the normalization factors $\frac{e^{i\theta\,\eta_3/2}}{\sigma\left(i\theta/2 \right)}$ have been chosen for convenience.

Given this form of $\Delta(x)$, since $\zeta(x)\equiv d/dx \ln \sigma(x)$, we find
\begin{equation}
\Delta^\prime=A\left\{ \zeta(A x+i{\bf K}^\prime-i\theta/2)-\zeta(A x+i{\bf K}^\prime)+i\lambda/A\right\}\Delta
\end{equation}
and since ${\mathcal P}\equiv-\zeta^\prime$, we find
\begin{equation}
\Delta^{\prime\prime}=A^2\left\{ \left[-{\mathcal P}(A x+i{\bf K}^\prime-i\theta/2)+{\mathcal P}(A x+i {\bf K}^\prime)\right]+ \left[\zeta\left(A x+i{\bf K}^\prime-i\theta/2\right)-\zeta\left(A x+i{\bf K}^\prime\right)+i \lambda/A\right]^2\right\}\Delta
\end{equation}
Furthermore, using the quasiperiodicity properties of the Weierstrass functions (\ref{sigmaperiod})
and the Weierstrass function product formula (\ref{addition1}), it follows that
\begin{eqnarray}
|\Delta(x)|^2&=&A^2 \frac{\sigma\left(A x+i{\bf K}^{\prime}-i\theta/2\right)\sigma\left( A x+i{\bf K}^{\prime}+i\theta/2\right)}{\sigma^2\left(A x+i{\bf K}^\prime \right)\sigma\left(i\theta/2\right)\sigma\left(-i\theta/2\right)}\nonumber\\
&=& A^2\left[{\mathcal P}(A x+i{\bf K}^{\prime})-{\mathcal P}(i \theta/2)\right]
\label{mag}
\end{eqnarray}
Therefore,
\begin{eqnarray}
\Delta^{\prime\prime}-2|\Delta|^2 \Delta&=&A^2\left\{-\left[{\mathcal P}(A x+i{\bf K}^\prime-i\theta/2)+{\mathcal P}(A x+i {\bf K}^\prime)\right]\right.\nonumber\\
&&\left.+ \left[\zeta\left(A x+i{\bf K}^\prime-i\theta/2\right)-\zeta\left(A x+i{\bf K}^\prime\right)+i \lambda/A\right]^2  \right.\nonumber
\\
&&\left.
+ 2 {\mathcal P}(i \theta/2)  \right\}\Delta
\end{eqnarray}
where note that the relative sign between the first ${\mathcal P}$ functions has been flipped by the subtraction of $2|\Delta|^2\Delta$.

Now we use the remarkable identity (\ref{addition3}) that relates the ${\mathcal P}$ function to squares of the $\zeta$ function, to find that
\begin{equation}
\Delta^{\prime\prime}-2\Delta^2 \Delta=A^2\left[3{\mathcal P}(i\theta/2)+\left(\lambda/A+i\zeta(i\theta/2)\right)^2\right] \Delta+2iA\left[\lambda/A+i\zeta(i\theta/2)\right]\Delta^{\prime}
\end{equation}
Thus, $\Delta(x)$ of the form in (\ref{guess}) does indeed satisfy the NLSE (\ref{nlse}), and one simply needs to match the constants algebraically in order to express $a(E)$, $b(E)$ and ${\mathcal N}(E)$ in terms of $\theta$. This determines $\lambda$ in (\ref{guess}) up to an additive constant, which can be fixed by matching the fermion spectrum to the form in (\ref{band-edges}).

\subsection{Reduction of general solution to special cases}
\label{reductions}

To conclude this section describing our new complex crystalline solution (\ref{complex-kink-crystal}), we note that it incorporates all previously known solutions as special cases.

\subsubsection{Reduction to real kink crystal condensate}

When $\theta=2{\bf K}^\prime$, the scale factor $A\to 2m/(1+\sqrt{\nu})$,  and the condensate reduces to the real kink crystal in (\ref{real-kink-crystal}):
\begin{eqnarray}
\Delta &\to & \frac{2m \sqrt{\nu}}{1+\sqrt{\nu}}\, {\rm sn}\left(\frac{2m \, x}{1+\sqrt{\nu}}; \nu\right) \nonumber\\
&=& m\, \tilde{\nu}\, \frac{{\rm sn}\left(m\, x; \tilde{\nu}\right){\rm cn}\left(m\, x; \tilde{\nu}\right)}{{\rm dn}\left(m\, x; \tilde{\nu}\right)}\quad , \quad \tilde{\nu}=\frac{4\sqrt{\nu}}{(1+\sqrt{\nu})^2} \quad
\end{eqnarray}
In this limit, the amplitude vanishes at $x=0$, and one sees that the kink "winds" through angle $2\varphi=-\pi$ by passing through zero. For other values of $\theta$, the complex kink crystal winds around zero, but without the amplitude actually vanishing. For $\theta=2{\bf K}^\prime$, the bound band becomes symmetric about $E=0$, because the band edges (\ref{band-edges}) reduce to:
\begin{eqnarray}
E_2&\to&-m \left(\frac{1-\sqrt{\nu}}{1+\sqrt{\nu}}\right) \nonumber\\
E_3&\to&m \left(\frac{1-\sqrt{\nu}}{1+\sqrt{\nu}}\right)
\end{eqnarray}
Accordingly, the resolvent functions $a(E)$, $b(E)$ and ${\mathcal N}(E)$ in (\ref{complex-kink-crystal-res}) reduce smoothly when $\theta=2{\bf K}^\prime$ to the corresponding expressions for the real kink crystal in (\ref{real-kink-crystal-res}).

\subsubsection{Reduction to single complex kink condensate}

For general $\theta$, when $\nu\to1$,  the complex crystalline condensate in (\ref{complex-kink-crystal})  reduces to  Shei's complex kink  solution (\ref{complex-kink}). To see this, first observe that in this limit, $A(m, \theta, \nu)\to m \, \sin(\theta/2)$, and ${\bf K}$ diverges, while ${\bf K}^\prime\to \pi/2$. Thus the period $2{\bf K}/A$ diverges. Also, the band edges contract as $E_2\to E_3=m \cos(\theta/2)$, so the band shrinks to a single bound state, whose $\theta$ dependence matches that of the bound state for Shei's complex kink condensate.
Furthermore, as $\nu\to 1$, the Weierstrass functions simplify:
\begin{eqnarray}
\sigma(x)&\to& \sinh(x)e^{-x^2/6} \nonumber\\
\zeta(x)&\to& -\frac{x}{3}+\coth(x) \nonumber\\
{\mathcal P}(x)&\to& \frac{1}{3}+\frac{1}{\sinh^2(x)}
\label{w1}
\end{eqnarray}
These relations show that on the interval $[-{\bf K}/A, {\bf K}/A]$, in the limit $\nu\to 1$, the complex crystalline condensate (\ref{complex-kink-crystal}) reduces to Shei's single complex kink condensate (\ref{complex-kink}) up to an unimportant constant phase factor.

\subsubsection{Reduction to single plane wave condensate}

In the opposite limit, as $\nu\to 0$, the period remains finite, because ${\bf K}\to \pi/2$, and $A\to 2\tanh(\theta/4)$. As $\nu\to 0$, the Weierstrass functions simplify:
\begin{eqnarray}
\sigma(x)&\to& \sin(x)e^{x^2/6} \nonumber\\
\zeta(x)&\to& \frac{x}{3}+\cot(x) \nonumber\\
{\mathcal P}(x)&\to& -\frac{1}{3}+\frac{1}{\sin^2(x)}
\label{w0}
\end{eqnarray}
Thus, the complex crystalline condensate (\ref{complex-kink-crystal}) reduces to a single plane wave (chiral spiral) form:
\begin{equation}
\Delta \to m\,  {\rm sech}^2(\theta/4) e^{-2im\, \tanh^2(\theta/4) \, x}
\label{nu0}
\end{equation}

\section{Solutions to the Bogoliubov/de Gennes equation
\label{solutions}}

\subsection{Spinor solutions}
\label{spinors}

Remarkably, not only is it possible to solve the NLSE (\ref{nlse}) exactly, we can also solve exactly the associated BdG equation \cite{smirnov}. In the previous sections we described the spectrum;  here we present the explicit spinor solutions and express the spectral information in a more compact and useful form. We write the BdG equation  (\ref{dbdg}) as
\begin{equation}
\begin{pmatrix}
{-i\frac{d}{dx}&\Delta(x)\cr \Delta^*(x) & i\frac{d}{dx}}
\end{pmatrix}\psi=E\psi
\label{dbdg2}
\end{equation}
The solutions for the real condensates in Section \ref{real} are well known \cite{dhn,feinberg}. For the complex plane wave, $\Delta(x)=A e^{iqx}$, the solutions are simply chiral rotations of free spinors, and are discussed for example in \cite{ohwa}. For Shei's complex kink condensate (\ref{complex-kink}), the spinor solutions are given in \cite{shei,thies-gl}. For the complex kink crystal (\ref{complex-kink-crystal}), the two independent spinor solutions can be written as 
\begin{eqnarray}
 \psi_{-}&= & \sqrt{\frac{A}{2| dk/d\alpha |}} e^{-i\eta_{3} \alpha}
\left( \begin{array}{c}
\mbox{$\frac{\sigma(Ax+i{\bf K}^\prime+i\alpha-i\theta/4)}{\sigma(Ax+i{\bf K}^\prime)\, \sigma(i\alpha-i\theta/4)} e^{\frac{iAx}{2} [-i\zeta(i\theta/2)+i\,{\rm ns}(i\theta/2)]+i\theta \eta_{3}/4+i\pi/4}$}
 \\
 \\
\mbox{$\frac{\sigma(Ax+i{\bf K}^\prime+i\alpha+i\theta/4)}{\sigma(Ax+i{\bf K}^\prime)\, \sigma(i\alpha+i\theta/4)} e^{-\frac{iAx}{2} [-i\zeta(i\theta/2)+i\,{\rm ns}(i\theta/2)]-i\theta \eta_{3}/4-i\pi/4}$}
\end{array}\right)e^{\frac{iAx}{2} [i\zeta(i\alpha+i\theta/4)+i\zeta(i\alpha-i\theta/4)]}  \nonumber
\\
\label{wave}
\\
\psi_{+}&=&-  \sqrt{\frac{A}{2| dk/d\alpha |}}e^{i\eta_{3} \alpha}
\left( \begin{array}{l}
\mbox{$\frac{\sigma(Ax+i{\bf K}^\prime-i\alpha-i\theta/4)}{\sigma(Ax+i{\bf K}^\prime)\,\sigma(i\alpha+i\theta/4)} e^{\frac{iAx}{2} [-i\zeta(i\theta/2)+i\,{\rm ns}(i\theta/2)]+i\theta \eta_{3}/4+i\pi/4}$}
 \\
 \\
\mbox{$\frac{\sigma(Ax+i{\bf K}^\prime-i\alpha+i\theta/4)}{\sigma(Ax+i{\bf K}^\prime)\,\sigma(i\alpha-i\theta/4)} e^{-\frac{iAx}{2} [-i\zeta(i\theta/2)+i\,{\rm ns}(i\theta/2)]-i\theta \eta_{3}/4-i\pi/4}$}
 \end{array}\right)e^{-\frac{iAx}{2} [i\zeta(i\alpha+i\theta/4)+i\zeta(i\alpha-i\theta/4)]} \nonumber
\end{eqnarray}
Here, $\alpha$ is a spectral parameter that characterizes the energy and momentum of these solutions, and $k$ is the momentum, defined below in (\ref{bloch-momentum}). 
Notice that $\psi_\pm$ differ from one another by the sign of $\alpha$.

To relate the energy  eigenvalue $E$ to the spectral parameter $\alpha$, we substitute these forms into the BdG equation (\ref{dbdg2}), and make use of the Weierstrass function identity (\ref{addition2}). This identity immediately shows that these are indeed solutions, and determines the energy to be
\begin{eqnarray}
E(\alpha)&=& \frac{A}{2}
[i\,\zeta(i\alpha-i\theta/4)-i\,\zeta(i\alpha+i\theta/4)+i\,\zeta(i\theta/2)+i\,{\rm ns}(i\theta/2)]\nonumber\\
&=& -m+2m\left(\frac{{\mathcal P}(i\theta/4)-{\mathcal P}(i{\bf K}^\prime)}{{\mathcal P}(i\theta/4)-{\mathcal P}(i\alpha)}\right)
\label{en}
\end{eqnarray}
where $A$ is given by (\ref{factor}).
\begin{figure}
\includegraphics[scale=0.6]{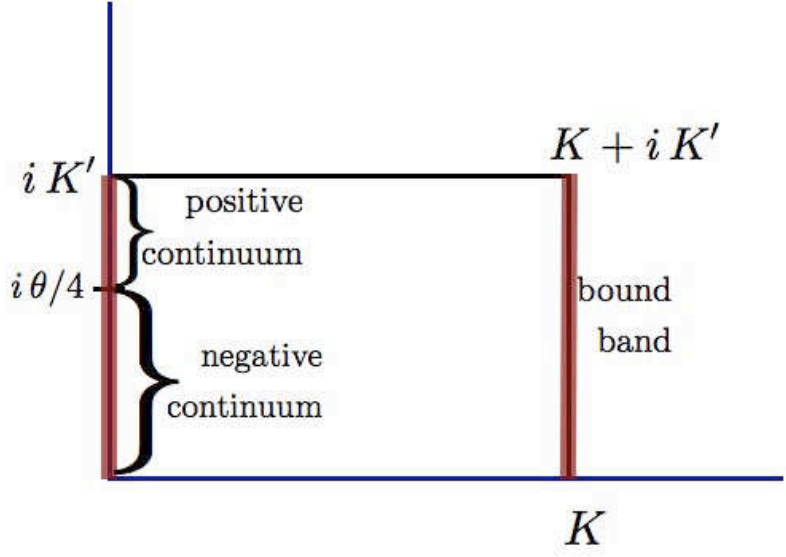}
\caption{The fundamental rectangle for the spectral parameter $i\alpha$ appearing in the spinor solutions (\ref{wave}) to the BdG equation. The bound band is characterized by  ${\bf K}(\nu)\leq i\alpha\leq {\bf K}(\nu)+i{\bf K}^\prime(\nu)$, while the positive and negative energy continua are characterized by $0\leq i\alpha\leq i{\bf K}^\prime(\nu)$. The point $\alpha=\theta/4$ represents $E=\pm\infty$, depending on the side of approach.}
\label{fig10}
\end{figure}
The spectral parameter $\alpha$ is defined on the vertical edges,  $0\leq i\alpha\leq i{\bf K}^\prime(\nu)$,  ${\bf K}(\nu)\leq i\alpha\leq {\bf K}(\nu)+i{\bf K}^\prime(\nu)$, of the fundamental rectangle shown in Figure \ref{fig10}. Only on these edges of the fundamental rectangle are the spinor wavefunctions in (\ref{wave}) bounded, and is the energy $E(\alpha)$  real. The vertices of the rectangle correspond to the band edges [compare with equation (\ref{band-edges})]:
\begin{eqnarray}
E_1&=&E(i\alpha=0)=-m\nonumber\\
E_2&=&E(i\alpha={\bf K})=m(-1+2\, {\rm nc}^2(i\theta/4)) \nonumber\\
E_3&=&E(i\alpha={\bf K}+i{\bf K}^\prime)=m(-1+2\, {\rm nd}^2(i\theta/4))\nonumber \\
E_4&=&E(i\alpha=i{\bf K}^\prime)=+m
\label{band-edges-2}
\end{eqnarray}
\begin{figure}
\includegraphics[scale=0.6]{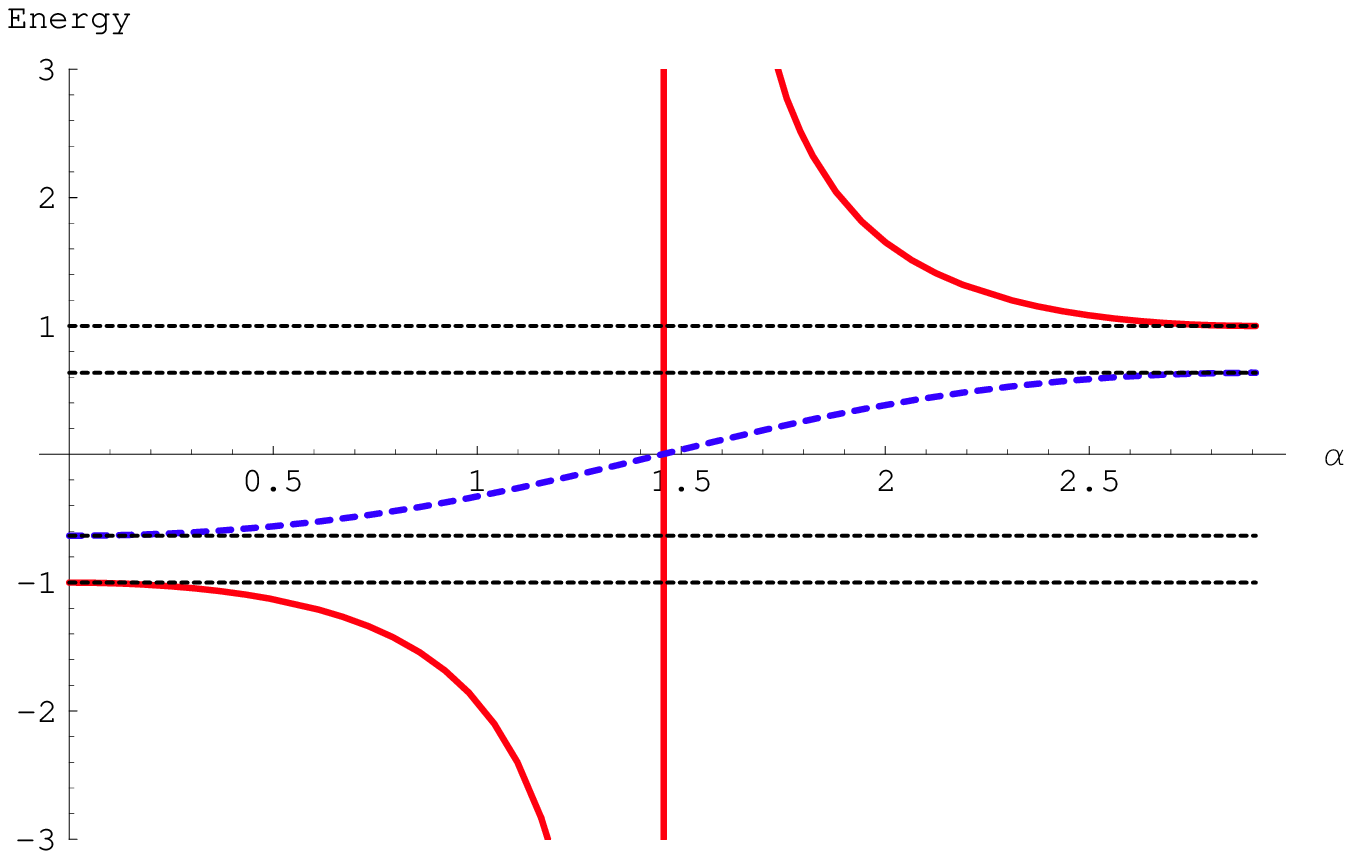}
\includegraphics[scale=0.6]{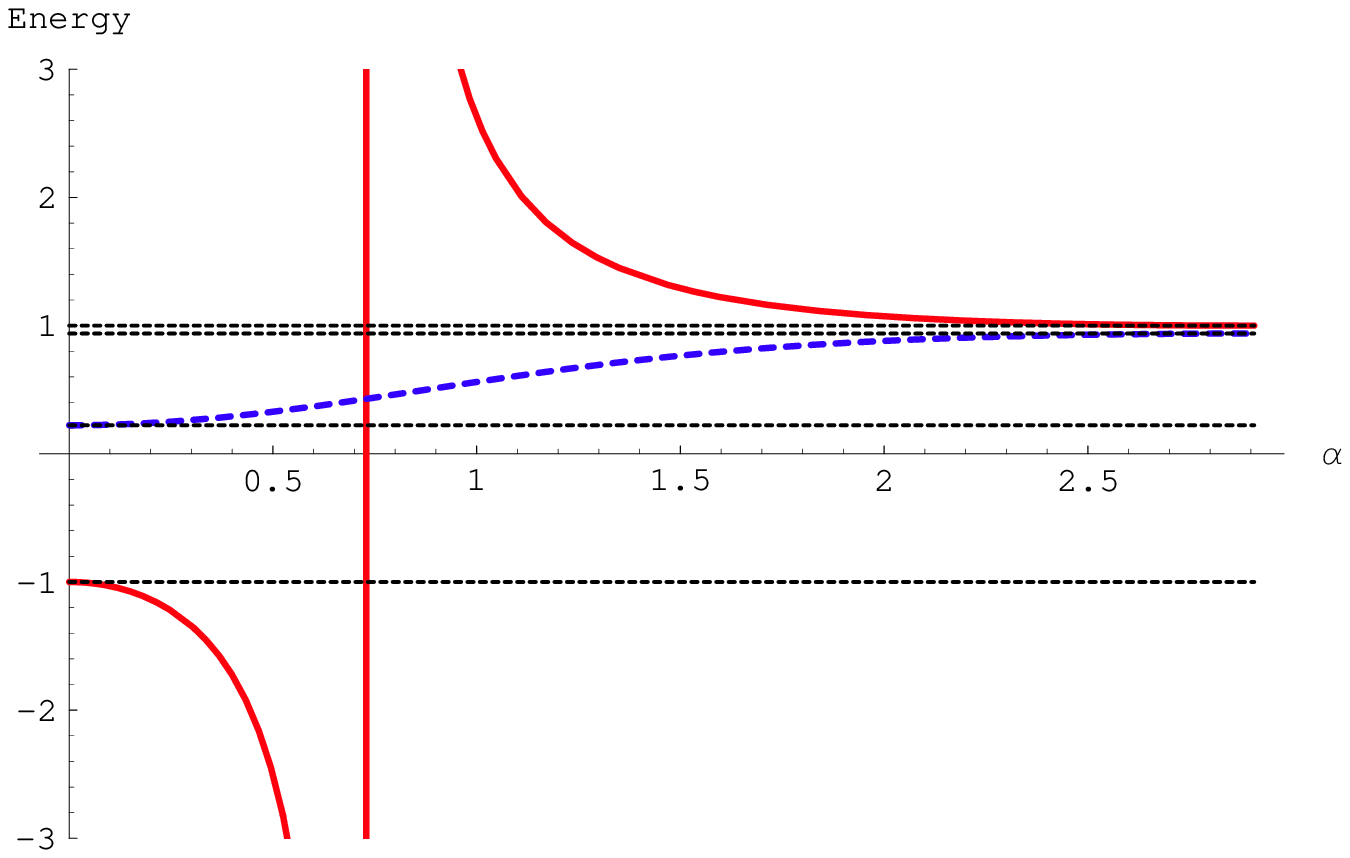}
\caption{Energy $E(\alpha)$ from (\ref{en}) as a function of the spectral parameter $\alpha$, for two different values of $\theta$. The solid [red] curves show the continuum energies, while the dashed [blue] curves show the energy in the bound band. The horizontal dashed lines denote the band edge energies $E_1$, $E_2$, $E_3$ and $E_4$ from (\ref{band-edges-2}), and the vertical [red] line gives the asymptote to $E=\mp\infty$ at $\alpha=\theta/4$, as discussed in the text. These curves are for elliptic parameter $\nu=0.05$. The first plot is for $\theta=2{\bf K}^\prime=5.82$, so that the band is symmetric about the origin. The second plot is for $\theta={\bf K}^\prime=2.91$, so the band is asymmetrically offset from the origin.}
\label{fig11}
\end{figure}
The right-hand boundary of the fundamental rectangle corresponds to the bound band, while the left-hand boundary corresponds to the positive and negative energy continua. The point $\alpha=\theta/4$ is associated with the point at infinity; the bottom of the Dirac sea is approached as $\alpha\to \theta/4$ from below, and the top of the positive energy continuum as $\alpha\to \theta/4$ from above. This is depicted in Figure \ref{fig11}.

To identify the momentum associated with these solutions, we recall that the quasiperiodic winding (\ref{winding-1}) of the condensate in (\ref{complex-kink-crystal}) implies that the BdG hamiltonian (\ref{ham-bdg}) is invariant under a period shift, up to a global chiral rotation through the winding angle $\varphi$:
\begin{eqnarray}
H\left(x+L\right)&=&e^{i\gamma_5 \varphi}\, H\left(x\right)\, e^{-i\gamma_5 \varphi}
\label{rotation}
\end{eqnarray}
Using the quasiperiodicity properties of the Weierstrass functions (\ref{sigmaperiod}), we see that under a period shift, the spinors in (\ref{wave}) acquire a chiral rotation and a Bloch phase:
\begin{eqnarray}
\psi_{\pm}(x+L)&=&e^{\pm i k L} e^{i\varphi\gamma_5} \psi_\pm(x)
\label{bloch-spinor}
\end{eqnarray}
where $\varphi$ is the winding angle defined in (\ref{winding-2}).
The Bloch momentum $k$ is expressed in terms of the spectral parameter $\alpha$ as:
\begin{eqnarray}
 k(\alpha)=- \frac{A}{2}\left[i\,\zeta(i\alpha+i\theta/4)+i\,\zeta(i\alpha-i\theta/4) +2\eta\alpha/{\bf K}\right]
\label{bloch-momentum}
\end{eqnarray}
\begin{figure}
\includegraphics[scale=0.6]{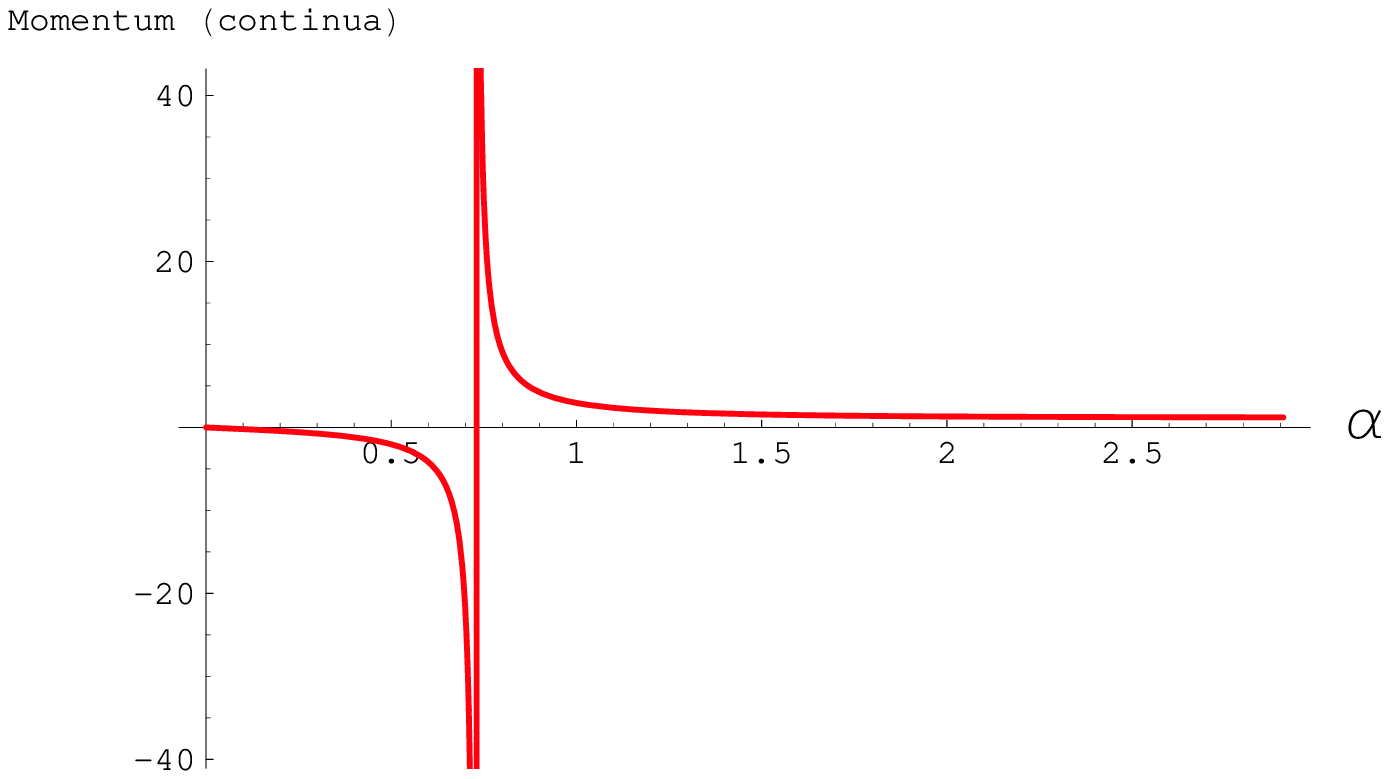}
\includegraphics[scale=0.6]{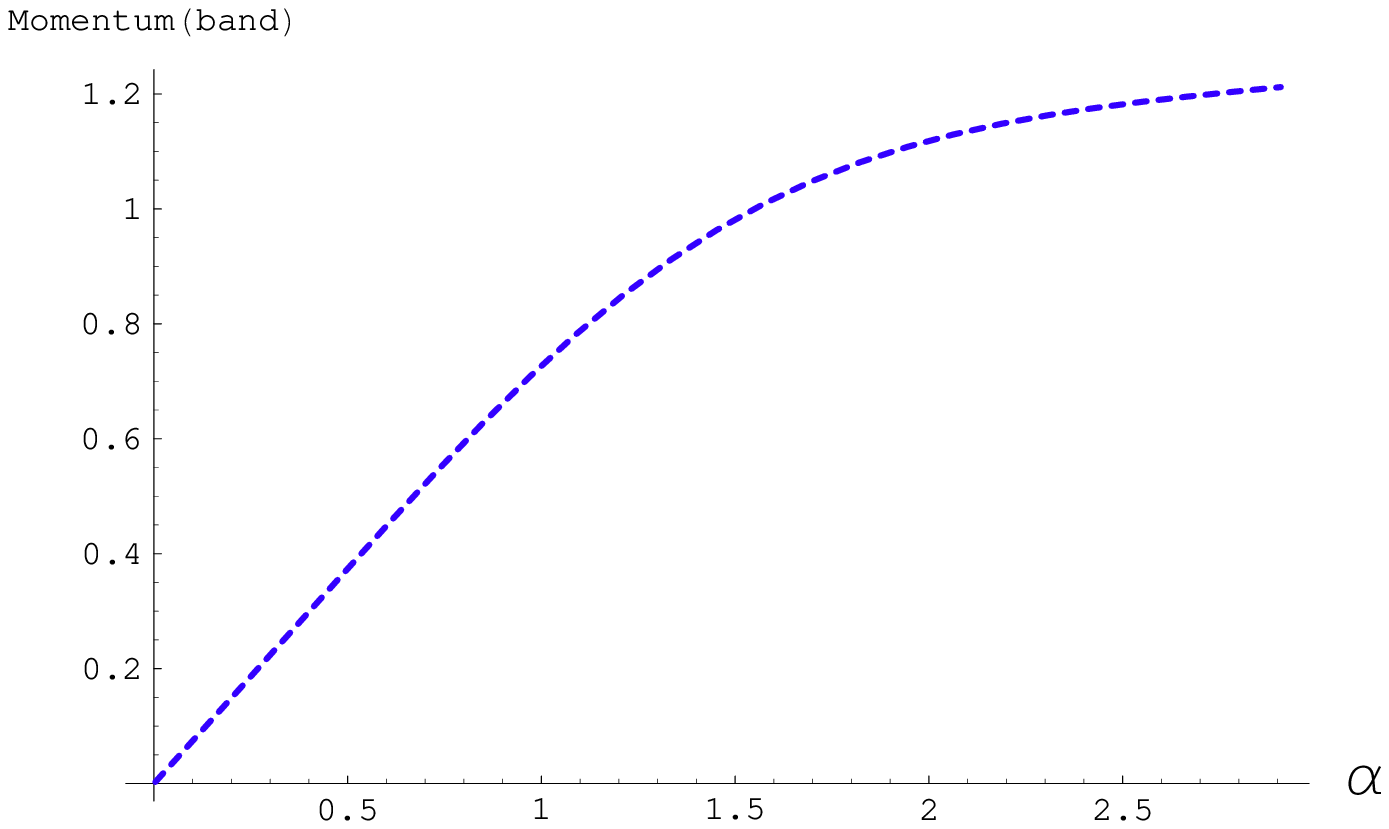}
\caption{Momentum as a function of the spectral parameter $\alpha$, in the positive and negative continua [first plot] and in the bound band [second plot]. Note that $dk/d\alpha$ is negative in the continuum and positive in the bound band. These plots are for $\nu=0.05$ and $\theta={\bf K}^\prime$, as in the second figure in Figure \ref{fig11}. The vertical [red] line in the first plot gives the asymptote to $k=\mp\infty$ at $\alpha=\theta/4$, as discussed in the text.}
\label{fig12}
\end{figure}
This is the relativistic version of Bloch's theorem. The momentum is real for $\alpha$ taking values on the vertical edges of the fundamental rectangle, and is plotted in Figure \ref{fig12}.
Since $k(\alpha)$ is odd in $\alpha$, we see that $\psi_\pm$ in (\ref{wave}) are positive and negative momentum solutions.

\subsection{Density of states}
\label{section-dos}

In this Section we present an efficient characterization of the density of states, and show how this relates to the resolvent in (\ref{ansatz}). In the previous section both the energy $E$ and the momentum $k$ were expressed in terms of the spectral parameter $\alpha$. Now consider the derivatives of the energy and momentum with respect to the spectral parameter $\alpha$. From (\ref{en}) and (\ref{bloch-momentum}) we see that
\begin{eqnarray}
\frac{dE}{d\alpha}&=&-\frac{A}{2}\left[{\mathcal P}\left(i \alpha+i\theta/4\right)-{\mathcal P}\left(i \alpha-i\theta/4\right)\right]\nonumber\\
\frac{dk}{d\alpha}&=&-\frac{A}{2}\left[{\mathcal P}\left(i \alpha+i\theta/4\right)+{\mathcal P}\left(i \alpha-i\theta/4\right)+2\eta/{\bf K}\right]
\label{alpha-derivative}
\end{eqnarray}
The significance of these expressions becomes clearer once we write the resolvent functions $a(E)$, $b(E)$ and ${\mathcal N}(E)$ in (\ref{complex-kink-crystal-res}) as functions of $\alpha$ (instead of $E$):
\begin{eqnarray}
a(\alpha)&=&\frac{A^2}{2}[2{\mathcal P}(i\theta/2)-{\mathcal P}(i\alpha+i\theta/4)-{\mathcal P}(i\alpha-i\theta/4)]
\\ b(\alpha)&=&A
[i\zeta(i\alpha-i\theta/4)-i\zeta(i\alpha+i\theta/4)+i\zeta(i\theta/2)-i{\rm ns}(i\theta/2)] \\
{\mathcal N}(\alpha)&=&\frac{\mp i}{A^2 [{\mathcal P}(i\alpha+i\theta/4)-{\mathcal P}(i\alpha-i\theta/4)]}
\label{complex-kink-crystal-res-alpha}
\end{eqnarray}
Here the upper sign in ${\mathcal N}$ is for the continuum states, and the lower sign is for the band. Thus, we recognize $dE/d\alpha$ as
\begin{eqnarray}
\frac{dE}{d\alpha}&=&\mp \frac{2}{A} \sqrt{(E^2-m^2)(E-E_2)(E-E_3)} \quad .
\label{ea}
\end{eqnarray}
And $dk/d\alpha$ can be written as
\begin{eqnarray}
\frac{dk}{d\alpha}&=&-A\left[{\mathcal P}\left(i\theta/2\right)- a(\alpha)/A^2+\eta/{\bf K}\right]\quad ,
\label{ka}
\end{eqnarray}
which is  negative in the continuum and positive in the band [see Figure \ref{fig12}].
Consequently, the density of states can be expressed as
\begin{eqnarray}
\frac{dk}{dE}&=& \frac{dk}{d\alpha}/\frac{dE}{d\alpha}\nonumber\\
&=&  \mp \frac{\left[a(E)-A^2{\mathcal P}(i\theta/2) -A^2 \eta/{\bf K}\right]}{2\sqrt{(E^2-m^2)(E-E_2)(E-E_3)}}
\label{dos}
\end{eqnarray}
When $\theta=2{\bf K}^\prime$, which is the ${\rm GN}_2$ limit [i.e., with a real condensate], this density of states reduces to
\begin{eqnarray}
\frac{dk}{dE}&=&  \mp \frac{E^2-m^2\frac{{\bf E}(\tilde{\nu})}{{\bf K}(\tilde{\nu})}} {\sqrt{(E^2-m^2)(E^2-m^2(1-\tilde{\nu}))}} \qquad  ,\qquad \tilde{\nu}\equiv\frac{4\sqrt{\nu}}{(1+\sqrt{\nu})^2}
\label{mom-real}
\end{eqnarray}
where ${\bf E}$ and ${\bf K}$ are the complete elliptic integrals \cite{as,ww}. This density of states (\ref{mom-real}) is precisely that found for the single-band finite-gap Schr\"odinger system of the ${\rm GN}_2$ model \cite{thies-gn}. The result in (\ref{dos}) generalizes this density of states to the general complex crystalline condensate (\ref{complex-kink-crystal}).

The density of states (\ref{dos}) has been derived from the energy and momentum of the spinor solutions in (\ref{wave}). For consistency, we compare this with the trace of the resolvent (\ref{ansatz}) over one period:
\begin{eqnarray}
\frac{1}{L}\int_{L}{\rm tr}_D\, R(x; E)dx&=&\frac{A}{2{\bf K}}\int_{-{\bf K}/A}^{+{\bf K}/A}2{\mathcal N}(E)\left(a(E)+|\Delta(x)|^2\right)dx\nonumber\\
&=& \frac{A\,{\mathcal N}(E)}{{\bf K}} \int_{-{\bf K}/A}^{+{\bf K}/A}dx\left[a(E)+A^2\left({\mathcal P}(A x+i {\bf K}^\prime)-{\mathcal P}(i\theta/2)\right)\right]\nonumber\\
&=& \frac{\left[a(E)-A^2{\mathcal P}(i\theta/2) - A^2 \eta/{\bf K}\right]}{2\sqrt{(m^2-E^2)(E-E_2)(E-E_3)}} \quad ,
\label{trace}
\end{eqnarray}
where we have used the integral
\begin{equation}
\int_{-{\bf K}/A}^{+{\bf K}/A}{\mathcal P}(A x+i {\bf K}^\prime)dx= -\frac{2\eta}{A} \quad .
\end{equation}
This  illustrates the consistency of our resolvent ansatz (\ref{ansatz}) with the spectral properties of the associated BdG equation.

\section{Hartree-Fock approach and the gap equation}
\label{hf}

So far we have shown that the gap equation (\ref{gap}) can be reduced to the NLSE (\ref{nlse}), and we have found the general solution to the NLSE. We now show the self-consistency of this approach by verifying that the gap equation in the form (\ref{gap-2}) is indeed satisfied. Inserting our ansatz form (\ref{ansatz}) for the resolvent, the gap equation in this form reads
\begin{eqnarray}
\Delta(x)=-2iN g^2 {\rm tr}_E\left[{\mathcal N}(E)\left(b(E) \Delta(x)-i\Delta^\prime(x)\right)\right]
\label{final-gap}
\end{eqnarray}
To verify (\ref{final-gap}), we require information about the single-particle fermionic spectrum of the BdG Hamiltonian for the condensate $\Delta(x)$ that we have found. This provides a complementary approach to finding a self-consistent condensate, by solving the associated relativistic Hartree-Fock problem.

\subsection{Solving the gap equation from the resolvent}
\label{gap-resolvent}

We first need to show that the coefficient of $\Delta^\prime(x)$ on the right-hand-side of  (\ref{final-gap}) vanishes once the energy trace is taken: 
\begin{eqnarray}
{\rm tr}_E\left[{\mathcal N}(E)\right]=0
\label{cond1}
\end{eqnarray}
This computation is greatly simplified by converting it into an integral over the spectral parameter $\alpha$. Recalling (\ref{ea}) and the normalization factor in (\ref{complex-kink-crystal-res}) we have
\begin{eqnarray}
{\mathcal N}(E)&=&\pm\frac{i}{2A} \frac{d\alpha}{dE} \quad .
\label{N2}
\end{eqnarray}
We assume that the negative energy continuum is fully occupied, and that the band inside the gap is partially occupied, by a fraction $\xi$. Then we can express (\ref{cond1}) as
\begin{eqnarray}
0 &=&\frac{i}{2\pi A} \left\{\int_0^{\theta/4} {d\alpha}
-\int_{-i{\bf K}}^{-i{\bf K}+\xi\, {\bf K}^\prime} {d\alpha}\right\}\nonumber\\
 &=&\frac{i}{2\pi A}(\theta/4-\xi\,{\bf K}^\prime)
\label{cond1-2}
\end{eqnarray}
Thus, $\theta$ must satisfy the condition
\begin{equation}
\frac{\theta}{4{\bf K}^\prime}=\xi = {\rm filling\,\, fraction}
\label{cond1-3}
\end{equation}
\begin{figure}
\includegraphics[scale=0.6]{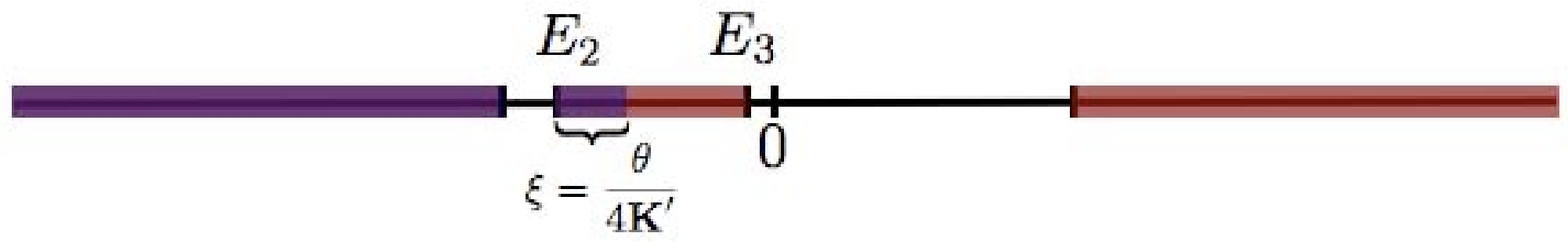}
\includegraphics[scale=0.6]{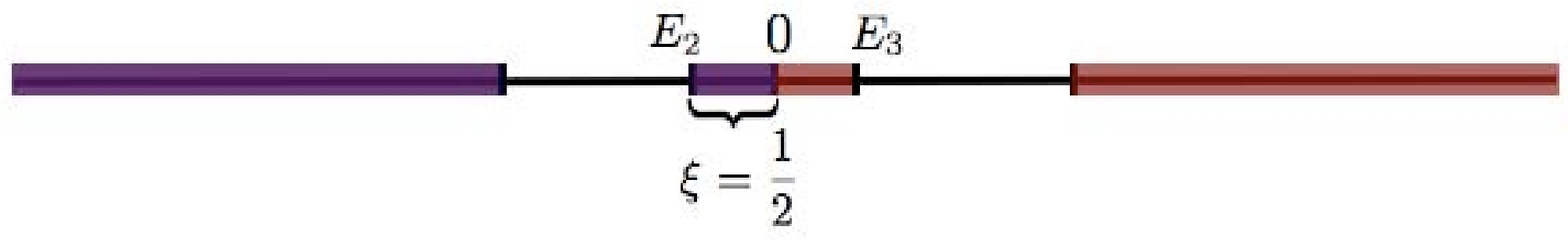}
\caption{The occupation of single-particle fermionic states for two different values of $\theta$. In general, to satisfy the gap equation the condition (\ref{cond1-3}) relates the condensate parameter $\theta$ to the filling fraction of the bound band. In the special case $\theta=2{\bf K}^\prime$, this filling fraction is $1/2$, and this corresponds o the real condensate of the discrete-chiral ${\rm GN}_2$ model, for which the spectrum is symmetric about 0. Thus, in this case, all the negative energy states are filled.}
\label{fig13}
\end{figure}

To appreciate this condition, we consider some special limiting cases. In the real [${\rm GN}_2$] limit, where $\theta=2{\bf K}^\prime$, this corresponds to filling the band half-way. Since the band is symmetrically placed about zero energy, this corresponds precisely to filling all the negative energy states, as in the conventional ${\rm GN}_2$ model analysis \cite{gross,dhn}. For the single kink, which is obtained in the infinite period limit, the bound state is half occupied \cite{jackiw,niemi}. 

For general $\theta$, in the infinite period limit, where ${\bf K}^\prime\to \pi/2$, we recover Shei's condition (\ref{shei-condition}) that $\theta/2\pi$ is the fractional filling of the bound state in the gap. Note that in this case, the fractional filling of this state is conventionally interpreted \cite{shei,fz} in terms of the filling of the level by a fixed fraction $\frac{n}{N}$ of the fermion flavors, with this fraction kept fixed in the infinite $N$ limit. At finite period, it is thermodynamically more natural to consider a fraction of the band being occupied by all flavors, as above. As the band contracts to a single bound state in the infinite period limit, we smoothly tend to the situation of having the level partially filled, while maintaining consistency with the gap equation for any period.

The other condition required for the gap equation (\ref{final-gap}) to be satisfied is that
\begin{eqnarray}
-2i N g^2 {\rm tr}_E\left[{\mathcal N}(E)\,b(E)\right]=1 \quad .
\label{cond2}
\end{eqnarray}
This is  the standard vacuum gap equation for the renormalization of the coupling \cite{gross,dhn,shei,fz}.

\subsection{Solving the gap equation from the spinor solutions}
\label{gap-spinors}

Given the spinor solutions of the BdG equation, we can reconstruct the expectation values $\langle\bar{\psi}(x)\psi(x)\rangle$ and $-\langle\bar{\psi}(x)i\gamma^5\psi(x)\rangle$, to verify that they correspond to the real and imaginary parts of condensate $\Delta(x)$. From (\ref{wave}), we find
\begin{eqnarray}
\bar{\psi}_k(x)\psi_k(x)-i\,\bar{\psi}_k(x) i \gamma^5\psi_k(x)&=&\psi^\dagger_k(x)(\gamma^0+\gamma^1)\psi_k(x)\nonumber\\
&&\hskip -3cm = \frac{iA}{|dk/d\alpha |}e^{-2i\eta_3\alpha} e^{iAx\left[-i\zeta(i\theta/2)+i{\rm ns}(i\theta/2)\right]}\frac{\sigma(Ax+i{\bf K}^\prime+i\alpha-i\theta/4)\sigma(Ax-i{\bf K}^\prime-i\alpha-i\theta/4)}{\sigma(Ax+i{\bf K}^\prime)\sigma(-i\alpha-i\theta/4)\sigma(i\alpha-i\theta/4)\sigma(Ax-i{\bf K}^\prime)}
\nonumber\\
&&\hskip -3cm = \frac{i}{|dk/d\alpha |}\left[\zeta(Ax+i{\bf K}^\prime)+\zeta(i\alpha-i\theta/4)-\zeta(Ax+i{\bf K}^\prime-i\theta/2)-\zeta(i\alpha+i\theta/4)\right]\,\Delta(x)
\nonumber\\
&&\hskip -3cm =  \frac{1}{|dk/d\alpha |}\frac{1}{A}\left[b(E)\,\Delta(x)-i\Delta^\prime(x)\right]
\label{HFgap}
\end{eqnarray}
where the subscript $k$ on $\psi_k$ emphasizes that these are spinor solutions of a given momentum $k$. In deriving (\ref{HFgap})  we have used the product identity (\ref{addition2}) for the Weierstrass sigma function.

Thus, for a given momentum $k$, $\bar{\psi}_k(x)\psi_k(x)-i\,\bar{\psi}_k(x) i \gamma^5\psi_k(x)$ has the same $x$ dependence as the upper off-diagonal entry of the resolvent, which appears in the gap equation (\ref{final-gap}). To compute the expectation value we trace over the $N$ flavors and integrate over momentum:
\begin{eqnarray}
\langle \bar{\psi}(x)\psi(x)\rangle -i \langle \bar{\psi}(x)i\gamma^5 \psi(x)\rangle&=&N\int \frac{dk}{2\pi} \left[
\bar{\psi}_k(x)\psi_k(x)-i\,\bar{\psi}_k(x) i \gamma^5\psi_k(x)\right]\nonumber\\
&=& \frac{N}{2\pi A}\int d\alpha\, {\rm sgn}\left(\frac{dk}{d\alpha}\right) \left[b(\alpha)\Delta(x)-i\Delta^\prime(x)\right]
\end{eqnarray}
Thus we have precisely the same integrals as were evaluated in the previous section to show that the gap equation is satisfied.
The vanishing of the coefficient of $\Delta^\prime(x)$ leads to the same relation (\ref{cond1-3}) between the filling fraction and $\theta$, while the the coefficient of $\Delta(x)$ is again the vacuum gap equation (\ref{cond2}) expressing the renormalization of the coupling constant g.

It is interesting to note that while the condensate $\Delta(x)$ is spatially inhomogeneous, the expectation value of the charge density is spatially uniform. From the explicit spinor solutions in (\ref{wave}) we find:
\begin{eqnarray}
\psi^\dagger_k(x)\psi_k(x)= \frac{1}{| dk/d\alpha |}\frac{1}{A}\left(|\Delta(x)|^2+a(\alpha)\right)
\label{charge}
\end{eqnarray}
For a given single particle state of momentum $k$, the charge density $\psi^\dagger_k(x)\psi_k(x)$ is spatially inhomogeneous, but this inhomogeneity is washed out by integrating over $k$, so that the expectation value is uniform:
\begin{eqnarray}
\frac{d}{dx} \langle \psi^\dagger(x)\psi(x)\rangle=0
\label{constant-charge}
\end{eqnarray}
Indeed, for a given $k$, the charge density $\psi^\dagger_k(x)\psi_k(x)$ has spatial inhomogeneity determined by $|\Delta(x)|^2$, the magnitude squared of the condensate. But the coefficient of this $x$-dependent term in (\ref{charge}) is precisely the same as the coefficient of the $\Delta^\prime(x)$ term in (\ref{HFgap}). We just saw above that this term vanishes [when the energy trace is taken], in order to satisfy the gap equation. Thus, the very same condition (\ref{cond1-3}) that ensures that the gap equation is satisfied also shows that $\langle\psi^\dagger(x)\psi(x)\rangle$ is uniform. 

There is another, more physical, way to understand this fact \cite{karbstein}.
It expresses the conservation of axial charge in the ${\rm NJL}_2$ model. Recall that the ${\rm NJL}_2$ system has two conserved currents: the first is the charge current $j^\mu=\bar{\psi}\gamma^\mu \psi$, and the second is the axial current $j^\mu_5=\bar{\psi}\gamma^\mu \gamma^5 \psi$. In $1+1$ dimensions these are related by $j_5^\mu =\epsilon^{\mu\nu} j_\nu$. For static but spatially inhomogeneous condensates, the charge current conservation
is automatically satisfied. 
Axial current conservation is more interesting. For a static condensate,
\begin{eqnarray}
\frac{\partial}{\partial x^\mu}\langle j^\mu_5(x)\rangle=\frac{d}{dx} \langle \psi^\dagger(x)\psi(x)\rangle
\end{eqnarray}
so axial charge conservation requires a uniform charge expectation value, as was found above (\ref{constant-charge}) from the single-particle spectral properties. Another way to see this is to observe that we can express the expectation value of the axial current in terms of the resolvent
\begin{eqnarray}
\langle j^\mu_5 \rangle&=&tr_{E, D} \left(\gamma^0 \gamma^\mu \gamma^5 R(x; E)\right)
\end{eqnarray}
Taking the spatial derivative and using the Eilenberger equation (\ref{dikii}), we find
\begin{eqnarray}
\frac{d}{dx}  \langle j^1_5 \rangle&=&tr_{E, D} \left(\partial_x R(x; E)\right)\nonumber\\
&=& 2\sigma(x) \langle \bar{\psi}(x)i \gamma^5 \psi(x)\rangle-2 \pi(x) \langle \bar{\psi}(x) \psi(x)\rangle
\end{eqnarray}
Thus, once again, the gap equation ensures that axial current conservation is satisfied. Interestingly, it is encoded into the Eilenberger equation (\ref{dikii}).

On the other hand, in the ${\rm GN}_2$ model, which has just a discrete chiral symmetry, there is no axial current conservation and  $\pi(x)=0$. Then the Eilenberger equation (\ref{dikii}) again expresses the correct relation
\begin{eqnarray}
\partial_\mu \langle j^\mu_5\rangle=2\sigma(x)\langle \bar{\psi}(x)i\gamma^5\psi(x)\rangle
\end{eqnarray}
and  $\langle \bar{\psi}(x)\psi(x)\rangle$, $\sigma(x)$ and $\langle \bar{\psi}(x)i\gamma^5\psi(x)\rangle$are all inhomogeneous.

\section{All-Orders Ginzburg-Landau Expansion}
\label{allorders}

In this Section we present another perspective on our solution, in order to illustrate what aspects might possibly be extended to higher dimensional inhomogeneities. While a direct application of the resolvent approach, or the inverse scattering approach, to higher dimensions is problematic, one approach that can be straightforwardly generalized to higher dimensions consists of the Ginzburg-Landau expansion of the effective action. In this approach, one expands the free energy corresponding to the effective action (\ref{effective}) in powers of the static condensate $\Delta$ [i.e, the order parameter] and its spatial derivatives. Using the relation (\ref{spectral}) between the spectral function $\rho(E)$ and the trace of the resolvent, we write the free energy density (per flavor) as
\begin{equation}
{\mathcal E}_{\rm free}(x)=\frac{1}{\pi \beta} \int_{-\infty}^\infty  dE\,{\rm Im}\left[{\rm tr}_{D}\, R(x; E+i\epsilon)\right] \ln\left(1+e^{-\beta(E-\mu)}\right)
\label{free}
\end{equation}
Thus, knowing the resolvent means we know the free energy. Usually, it is impossible to evaluate this local density of states exactly. However, an asymptotic expansion of $R(x; E)$ can always be obtained from a Laplace transform of the heat kernel expansion, leading to
\begin{eqnarray}
R(x; E)=\frac{1}{2}\sum_{n=0}^\infty\frac{r_n(x)}{E^n}
\label{r-expansion}
\end{eqnarray}
We stress that such an expansion can be derived in any dimension, not just in $1+1$ dimensions. This is because the heat kernel expansion is known in any dimension. However, in $1+1$ dimensions we now have the luxury of having found an exact solution. In this Section we compare this exact solution with the Ginzburg-Landau expansion, and show in explicit detail how the gap equation for the inhomogeneous condensate is solved order-by-order in this Ginzburg-Landau expansion. The way this works is surprisingly simple and elegant, as shown below.

In $1+1$ dimensions, including the Hartree-Fock double-counting correction, and renormalizing, leads to the standard Ginzburg-Landau expansion of the effective Lagrangian,  the low order terms of which are
\begin{eqnarray}
{\mathcal L}_{\rm GL}&=&\alpha_0+\alpha_2|\Delta|^2 
+\alpha_3 {\rm Im}\left[\Delta (\Delta^\prime)^*\right] 
+\alpha_4\left[|\Delta|^4+|\Delta^\prime |^2\right]
+\alpha_5{\rm Im}\left[ \left(\Delta^{\prime\prime}-3|\Delta |^2\Delta\right)(\Delta^\prime)^*\right]  \nonumber \\
&&
+\alpha_6\left[ 2|\Delta |^6+8 |\Delta |^2 |\Delta^\prime |^2+2 {\rm Re}\left((\Delta^\prime)^2(\Delta^*)^2\right)
+|\Delta^{\prime\prime}|^2\right]+\dots
\label{gl}
\end{eqnarray}
The low order terms are relatively simple, but high orders rapidly become cumbersome. The coefficients $\alpha_n(T, \mu)$ are known functions of temperature and chemical potential. For example, in $1+1$ dimensions \cite{thies-gl,buzdin3}
\begin{eqnarray}
\alpha_0&=&-\frac{\pi^2T}{6}-\frac{\mu^2}{2\pi}\nonumber\\
\alpha_2&=&\frac{1}{2\pi}\left[ \ln(4\pi T)+{\rm Re}\,\psi\left(\frac{1}{2}+i\frac{\beta\mu}{2\pi}\right)\right]\nonumber\\
\alpha_3&=& -\frac{1}{2^3\pi^2 T}{\rm Im}\,\psi^{(1)}\left(\frac{1}{2}+i\frac{\beta\mu}{2\pi}\right)\nonumber\\
\alpha_4&=&-\frac{1}{2^6\pi^3 T^2}{\rm Re}\,\psi^{(2)}\left(\frac{1}{2}+i\frac{\beta\mu}{2\pi}\right)\nonumber\\
\alpha_5&=&\frac{1}{2^8\pi^4 3\, T^3}{\rm Im}\,\psi^{(3)}\left(\frac{1}{2}+i\frac{\beta\mu}{2\pi}\right)\nonumber\\
\alpha_6&=&\frac{1}{2^12\pi^5 3\, T^4}{\rm Re}\,\psi^{(4)}\left(\frac{1}{2}+i\frac{\beta\mu}{2\pi}\right)
\label{alphas}
\end{eqnarray}
Here $\psi^{(k)}$ denotes the ${\rm k}^{\rm th}$ derivative of the Euler digamma function $\psi(z)=d \ln \Gamma(z)/dz$.

In higher dimensions, an analogous Ginzburg-Landau expansion can be derived. 
The form of the $\alpha_n(T, \mu)$ is different, but known, and the form of the spatial terms $r_n(x)$ is different, but computable. In this section we consider how the ${\rm NJL}_2$ model gap equation for inhomogeneous condensates looks in terms of the Ginzburg-Landau expansion.
We first present a simple recursive way to generate the expansion to all orders, and then we show how the gap equation is solved order by order.  A closely related expansion, the derivative [or gradient] expansion, which is an expansion just in powers of derivatives, but including all orders in powers of the condensate, is considered in \cite{stone}.

\subsection{Ginzburg-Landau expansion in $1+1$ dimension}
\label{2dgl}

The Ginzburg-Landau expansion of the effective action follows, by (\ref{logform}) and (\ref{spectral}), from an asymptotic expansion (\ref{r-expansion}) of the resolvent $R(x; E)$. 
The Eilenberger equation (\ref{dikii}) generates such an expansion recursively. It is more convenient to define (as in \cite{stone})
\begin{eqnarray}
g(x; E)\equiv R(x; E)\,\sigma_3
\label{g}
\end{eqnarray}
In terms of $g$, the Eilenberger equation (\ref{dikii}) reads
\begin{eqnarray}
g^\prime=iE[\sigma_3, g]-i[J, g] \quad , \quad J\equiv
\begin{pmatrix}
{0&-\Delta \cr
\Delta^*&0}
\end{pmatrix}
\label{g-dikii}
\end{eqnarray}
We now define an asymptotic expansion for $g$
\begin{eqnarray}
g(x; E)=-\frac{i}{2}\sum_{n=0}^\infty\frac{g_n(x)}{E^n}  \qquad ,\qquad  g_n(x)\equiv \begin{pmatrix}
{c_n(x)&-d_n(x) \cr
d_n^*(x)&-c_n(x)}
\end{pmatrix}
\label{g-expansion}
\end{eqnarray}
The factor $-i$ comes from the large $E$ expansion of ${\mathcal N}(E)$, as is already clear from the constant $\Delta$ case in (\ref{constant-resolvent}), which also tells us that $g_0=\sigma_3$. The Eilenberger equation implies the simple recursion formula for the $g_n(x)$:
\begin{eqnarray}
g_n^\prime(x)=i[\sigma_3, g_{n+1}(x)]-i[J(x), g_n(x)]
\label{recursion}
\end{eqnarray}
The determinant condition ${\rm det}\, g=\frac{1}{4}$ fixes the $c_n$ in terms of the $d_n$:
\begin{eqnarray}
c_0&=&1\nonumber\\
c_1&=&0\nonumber\\
c_2&=&\frac{1}{2}|d_1|^2\nonumber\\
c_3&=&\frac{1}{2}\left(d_2 d_1^*+d_1d_2^*\right)\nonumber\\
c_4&=&-\frac{1}{8}|d_1|^4+\frac{1}{2}|d_2|^2+\frac{1}{2}\left(d_3 d_1^*+d_1d_3^*\right)\nonumber\\
c_5&=&-\frac{1}{4}|d_1|^2\left(d_2 d_1^*+d_1d_2^*\right)+\frac{1}{2}\left(d_2 d_3^*+d_3d_2^*\right)+\frac{1}{2}\left(d_4 d_1^*+d_1d_4^*\right)\nonumber\\
&\vdots&
\label{c-d}
\end{eqnarray}
So, given $g_0=\sigma_3$, we learn from (\ref{c-d}) that $c_1=0$. Then the
 Eilenberger recursion equation (\ref{recursion}) determines $d_1=\Delta$, so that $g_1(x)=J(x)$. Next, knowing $d_1$, we learn from (\ref{c-d}) that $c_2=\frac{1}{2}|\Delta|^2$, and from the
 Eilenberger recursion equation (\ref{recursion}), we find that $d_2=-\frac{i}{2}\Delta^\prime$. Iterating this procedure we find:
\begin{eqnarray}
g_1(x)&=&J(x)\nonumber\\
g_2(x)&=&\frac{1}{2}\begin{pmatrix}
{|\Delta|^2 &i\Delta^\prime\cr
i\Delta^{\prime *}&-|\Delta|^2}
\end{pmatrix}
\nonumber\\
g_3(x)&=&-\frac{1}{4}\begin{pmatrix}
{i\left(\Delta^\prime \Delta^*-\Delta^{\prime *}\Delta\right) & -\left(\Delta^{\prime\prime} -2|\Delta|^2\Delta\right)\cr
\left(\Delta^{\prime\prime*} -2|\Delta|^2\Delta^*\right)& -i\left(\Delta^\prime \Delta^*-\Delta^{\prime *}\Delta\right)
}
\end{pmatrix}
\nonumber\\
g_4(x)&=& \frac{1}{8}\begin{pmatrix}
{3|\Delta|^4+3|\Delta^\prime |^2-\left(|\Delta |^2\right)^{\prime\prime} & -i\left(\Delta^{\prime\prime\prime}-6|\Delta |^2\Delta^\prime\right) \cr
-i\left(\Delta^{*\prime\prime\prime}-6|\Delta |^2\Delta^{*\prime}\right) & -3|\Delta|^4-3|\Delta^\prime |^2+
\left(|\Delta |^2\right)^{\prime\prime}
}
\end{pmatrix}\nonumber\\
&\vdots&
\end{eqnarray}
The spectral function is expressed in terms of the trace of the resolvent:
\begin{eqnarray}
{\rm tr}\, R(x; E)=-i\sum_{n=0}^\infty \frac{c_n(x)}{E^n}
\label{trace-exp}
\end{eqnarray}
Then the Ginzburg-Landau expansion (\ref{gl}) is obtained by inserting (\ref{trace-exp}) into (\ref{free}) and performing the energy integrals. Renormalization affects the first two terms, and the others follow from the integrals ($n>2$) \cite{konrad}:
\begin{eqnarray}
\frac{1}{\beta\pi} \int_{-\infty}^\infty dE\, {\rm Im}\left(\frac{-i}{(E+i\epsilon)^n}\right)
\log\left(1+e^{-\beta(E-\mu)}\right)&=&
\begin{cases}
{\frac{(-1)^{(n-2)/2} \beta^{n-2}}{(2\pi)^{n-1}(n-1)!}\, {\rm Re}\, \psi^{(n-2)}\left(\frac{1}{2}+i\frac{\beta\mu}{2\pi}\right) \quad, \quad {\rm n\,\, even}\cr\cr
\frac{(-1)^{(n-1)/2} \beta^{n-2}}{(2\pi)^{n-1}(n-1)!}\, {\rm Im}\, \psi^{(n-2)}\left(\frac{1}{2}+i\frac{\beta\mu}{2\pi}\right) \quad, \quad {\rm n\,\, odd}}
\end{cases}
\label{gl-full}
\end{eqnarray}

\subsection{Gap equation to all orders in Ginzburg-Landau expansion}
\label{gap-gl}

We now show how the gap equation (\ref{gap}) is solved order-by-order, to all orders in the Ginzburg-Landau expansion (\ref{gl-full}). There are two ways to see this. First,  even though the $c_n(x)$ at high order  $n$ become extremely complicated combinations of powers and derivatives of $\Delta$ and $\Delta^*$, when we evaluate $c_n(x)$ {\it on a  solution of the NLSE} (\ref{nlse}), they dramatically simplify and, for each $n$, reduce to something linear in $|\Delta(x)|^2$:
\begin{eqnarray}
\left[c_n(x)\right]_{NLSE}=\beta_n |\Delta(x)|^2+\gamma_n\quad , 
\label{fact-1}
\end{eqnarray}
for some constants $\beta_n$ and $\gamma_n$.
For example, for $c_2(x)=|\Delta(x)|^2$ it is obvious, and for $c_3(x)$ it follows immediately from (\ref{general-phase}):
\begin{eqnarray}
c_3(x)&=&-\frac{i}{4}\left(\Delta^\prime \Delta^*-\Delta^{\prime *}\Delta\right)\nonumber\\
&=& -\frac{1}{4}(b-2E)|\Delta |^2-{\rm constant}
\end{eqnarray}
For $c_4$, we note that the NLSE (\ref{nlse}) implies that $\left(|\Delta |^2\right)^{\prime\prime}=4|\Delta|^4+2|\Delta^\prime |^2+4\left(a-b^2/4-E^2\right)|\Delta|^2+{\rm constant}$, so that
\begin{eqnarray}
c_4(x)&\equiv&\frac{3}{8}\left(|\Delta|^4+|\Delta^\prime|^2-\frac{1}{3}\left(|\Delta|^2\right)^{\prime\prime} \right)
\nonumber\\
&=& \frac{1}{8}\left(-|\Delta|^4+|\Delta^\prime|^2-4\left(a-b^2/4-E^2\right)|\Delta|^2+{\rm constant} \right)
\end{eqnarray}
This expression is linear in $|\Delta |^2$ because we recall from (\ref{det-check}) and (\ref{general-phase}) that $\left(- |\Delta|^4+|\Delta^\prime|^2\right)$ is linear in $|\Delta|^2$; this ensures that  ${\rm det}\,R(x; E)=-\frac{1}{4}$ is satisfied, and shows that $c_4(x)$ reduces to the form in (\ref{fact-1}).
Indeed, (\ref{fact-1}) holds to {\it all orders} in the Ginzburg-Landau expansion --  the {\it diagonal} component of the Eilenberger equation (\ref{dikii}) is the generator of this infinite sequence of relations.

Another related fact is that when we seek stationary points of the Ginzburg-Landau effective action with respect to variation of the condensate, we generate a sequence of equations for the condensate. At higher order $n$, these equations become more and more complicated, involving higher derivatives and nonlinearities. But,
when evaluated on a solution to the NLSE, these variations take a much simpler form, for all $n$:
\begin{eqnarray}
\frac{\delta}{\delta \Delta^*(x)}\int dx\, c_n(x) =\delta_n \, \Delta(x)+\epsilon_n\,\Delta^\prime(x)
\label{fact-2}
\end{eqnarray}
for some constants $\delta_n$ and $\epsilon_n$.
It is straightforward but instructive to verify this remarkable property for low orders. To all orders, this infinite sequence of identities is generated by the {\it off-diagonal} entries of the Eilenberger equation (\ref{dikii}), so the off-diagonal entry of $R(x; E)$ in (\ref{ansatz}) explains why (\ref{fact-2}) holds at any order. To solve the gap equation, the parameters of the solution [i.e., the parameter $\theta$] must be adjusted so that the coefficient of $\Delta^\prime(x)$ vanishes, as explained in Section \ref{hf}. 

These observations may be a useful guide to seeking solutions to the gap equation in higher dimensions where neither inverse scattering nor the resolvent approach is directly applicable for periodic condensates, but where the Ginzburg-Landau expansion approach is  available.

\section{Conclusions}

To conclude, we have solved the inhomogeneous gap equation (\ref{gap}) in the ${\rm NJL}_2$ model, to obtain a self-consistent condensate (\ref{complex-kink-crystal}) that is crystalline in form. It is inherently complex, unlike the condensate in the discrete-chiral ${\rm GN}_2$ model where the condensate is real. The complex crystal has a periodic amplitude, while its phase rotates though a certain angle over each period. This complex crystalline condensate contains all other previously known solutions to the gap equation as special cases. In addition to finding the exact condensate, we have presented the exact solution to the associated Bogoliubov-de Gennes equation (\ref{dbdg}) which gives the spectrum of fermions quantized in the presence of such a condensate. The spinor wavefunctions and single-particle energy spectrum are found in closed form, and we confirmed the consistency of the results by solving the gap equation also by Hartree-Fock. The key technical idea in our approach is to use the form of the gap equation to motivate an ansatz (\ref{ansatz}) for the Gorkov resolvent. For any condensate, the resolvent must satisfy the Eilenberger-Dikii equation (\ref{dikii}), and so this reduces the problem to solving the nonlinear Schr\"odinger equation (\ref{nlse}). Given this solution, and in particular the exact density of states (\ref{dos}), we are now in a position to study exactly the phase diagram of the  ${\rm NJL}_2$ model, extending the Ginzburg-Landau analysis in \cite{bd1}. For example, once the free energy is minimized on our exact gap equation solution, the filling factor is then determined as a function of T and $\mu$. This will be addressed in future work.

An important possible extension of this work would be to the massive ${\rm NJL}_2$ model, in which a bare fermion mass is included in the original Lagrangian (\ref{lag}). This mass term explicitly breaks the continuous chiral symmetry, and approximate methods have found a rich structure in the associated phase diagram \cite{boehmer,thies-gl}. The resolvent approach presented here is easily extended to the massive case [yielding the real kink-antkink crystal condensate of the massive  ${\rm GN}_2$ system] only when the condensate is real. This is deeply related to the integrability properties of the classical equations of motion of these systems \cite{neveu}. More work is needed to understand analytically the situation relevant for the {\it massive} ${\rm NJL}_2$ system, where the condensate is complex. To date, no exact self-consistent inhomogeneous condensate solving the gap equation has been found for the massive ${\rm NJL}_2$ model, except for simple embeddings of the real solutions \cite{hillel}. Another possible extension is to study other $1+1$ dimensional models, such as the Schwinger model \cite{fischler,metlitski}.

The most interesting extension would be to try to extend some of these ideas to higher dimensions, for example for higher dimensional Gross-Neveu or Nambu-Jona Lasinio models \cite{hands}, or more ambitiously to search for crystalline condensates in QCD or QCD models 
\cite{klebanov,goldhaber,jackson,manton,bringoltz}. Here, the resolvent approach is not directly applicable, but the lessons from the way in which the Ginzburg-Landau expansion works (discussed in Section \ref{allorders}) may be useful. A numerical approach currently being developed  \cite{langfeld}, that does not rely on special features of $1+1$ dimensions, is a numerical evaluation of the free energy, using the worldline Monte Carlo approach of Gies and Langfeld \cite{gies}. Other new lattice methods have been developed recently \cite{gattringer,wolff2} to study four-fermion interactions, and it would be interesting to extend these to higher dimensions.

\section{Acknowledgments}
We thank the DOE for support through grant DE-FG02-92ER40716, and we thank Joshua Feinberg,  Dominik Nickel and Michael Thies for many helpful comments.

\section{Appendix A: Resolvent of a Dirac system, and the Eilenberger equation}

In this appendix we summarize the derivation of the Dikii-Eilenberger equation (\ref{dikii}) for the diagonal resolvent of a Dirac system. The interested reader is urged to consult also references \cite{waxman,dickey}, and Appendix B of the first paper in \cite{stone}. Here we sketch the key features of the argument, for the sake of being more self-contained. 

The important idea is simple. Recall that in one dimension, for a Schr\"odinger-like Sturm-Liouville operator, it is well known that the Green's function can be written in terms of a product of two independent solutions, normalized by their Wronskian \cite{ww}. An analogous construction exists for a one dimensional Dirac operator \cite{waxman,dickey,stone}. Furthermore, in the Sturm-Liouville case, the coincident-point limit $R(x,x; E)$ satisfies a differential equation, known as the Gel'fand-Dik'ii equation (for an excellent review, see \cite{feinberg}), just by virtue of being written as a product of solutions to the original differential equation \cite{ww}. Likewise, for a Dirac system, the coincident-point limit $R(x,x; E)$ also satisfies an equation, just by virtue of being expressed in terms of solutions to the original differential equation. This equation is the Dik'ii-Eilenberger equation (\ref{dikii}). The technical difference from the Gel'fand-Dik'ii equation arises because the Dirac operator is {\it first-order in derivatives} and because it is a $2\times 2$ {\it matrix operator}. 

So, consider two independent spinor solutions $\psi_{1,2}$ of the BdG equation $H\psi=E\psi$ in (\ref{dbdg}), where $H$ is the $2\times 2$ matrix first-order differential operator in (\ref{ham-bdg}). Then we can write the resolvent (also a $2\times 2$ matrix) as
\begin{eqnarray}
R(x, y; E)=
\begin{cases}
{\psi_1(x) R(y)\quad , \quad x<y \cr
\psi_2(x) L(y) \quad, \quad x>y}
\end{cases}
\end{eqnarray}
The row-vector functions $R(y)$ and $L(y)$ are determined by demanding that $R(x, y; E)$ satisfy
\begin{eqnarray}
\left(H-E\right)R(x, y; E)=\delta(x-y)
\end{eqnarray}
Integrating this first-order differential equation across the point $x=y$, we learn that 
\begin{eqnarray}
R(x, y; E)=\begin{cases}
{\frac{1}{iW} \psi_1(x) \psi_2^T(y)\,\sigma_1\quad , \quad x<y \cr
\frac{1}{iW}\psi_2(x) \psi_1^T(y)\,\sigma_1 \quad, \quad x>y}
\end{cases}
\end{eqnarray}
where the Wronskian $W$ is the scalar function $W(\psi_1,\psi_2)=i\psi_1^T\sigma_2\psi_2$. The coincident-point limit $x=y$ is given by an averaged limit as [see (\ref{product})]:
\begin{eqnarray}
R(x; E)=\frac{1}{2iW}\left(\psi_1(x)\psi_2^T(x)+\psi_2(x)\psi_1^T(x)\right)\sigma_1
\label{product-2}
\end{eqnarray}
Now, express the BdG equation (\ref{dbdg}) as
\begin{eqnarray}
\psi^\prime=i
\begin{pmatrix}
{E & -\Delta \cr
\Delta^* & -E}
\end{pmatrix}
\psi
\quad , \quad 
\psi^{\prime\,T}=i\psi^T
\begin{pmatrix}
{E & \Delta^* \cr
-\Delta & -E}
\end{pmatrix}
\label{dbdg-2}
\end{eqnarray}
Then, differentiating (\ref{product-2}) once with respect to $x$,  using the fact that each of $\psi_1$ and $\psi_2$ satisfies the BdG equation (\ref{dbdg}) in the form (\ref{dbdg-2}), we arrive immediately at the Dik'ii-Eilenberger equation (\ref{dikii}). Furthermore, if we write $\psi_1=(u_1, v_1)^T$, and $\psi_2=(u_2, v_2)^T$, then (\ref{product-2}) says that
\begin{eqnarray}
R(x; E)=\frac{1}{2i(u_1v_2-u_2 v_1)} 
\begin{pmatrix}
{u_1 v_2+u_2 v_1 & 2 u_1 u_2\cr
2 v_1 v_2 & u_1 v_2+ u_2 v_1}
\end{pmatrix}
\end{eqnarray}
Thus, the algebraic conditions (\ref{condition-1}) and (\ref{condition-2}) follow. The hermiticity condition (\ref{hermitean}) is less immediately obvious, because this depends on $E$ and $W$. However, it is clear from the definition of the resolvent that $R(x; E)=\langle x |1/(H-E) | x\rangle$ is hermitean for real $E$. The appropriate hermiticity properties are studied in more detail in \cite{waxman,stone,dickey}, and can also be seen from the result for the constant $\Delta$ case (\ref{constant-resolvent}), and the $i\,\epsilon$ condition used to define the spectral function (\ref{spectral}).

\section{Appendix B: Some Useful Properties of Elliptic Functions}

In this appendix we collect some basic facts and nontrivial identities for Weierstrass elliptic functions that are used repeatedly in this paper, in order to make the paper more self-contained. These functions play a special role because the self-consistent condensate $\Delta(x)$ and the spinor solutions $\psi(x)$ to the Bogoliubov-de Gennes equation are all expressed in terms of these functions. There are many good books on elliptic functions. Excellent classical references are \cite{as,ww}. We also found \cite{lawden,akhiezer} to be particularly useful. Very roughly speaking, the Weierstrass elliptic functions are doubly-periodic extensions of standard trigonometric functions:
\begin{eqnarray}
\sin(z) &\leftrightarrow& \sigma(z)\nonumber\\
\cot(z)=\frac{d}{dz}\ln \sin(z) &\leftrightarrow& \zeta(z)= \frac{d}{dz}\ln\sigma(z)\nonumber\\
\frac{1}{\sin^2(z)}=-\frac{d}{dz}\cot(z) &\leftrightarrow& {\mathcal P}(z)=-\frac{d}{dz}\zeta(z)
\label{weierstrass}
\end{eqnarray}
The trigonometric functions have periodicity properties along the real $z$ axis, but the Weierstrass functions are {\it doubly} (quasi-)periodic. They are specified by the real and imaginary (half-)periods, $\omega_1$ and $\omega_3$. It is standard to define also $\omega_2$ by $\omega_1+\omega_2+\omega_3=0$. Then the Weierstrass sigma function is quasi-periodic under shifts by $2\omega_i$:
\begin{equation}
\sigma(z+2 \omega_i)=-e^{2\eta_i(z+\omega_i)} \sigma(z)\quad , \quad i=1, 2, 3 \quad .
\label{sigmaperiod}
\end{equation}
Here $\eta_i\equiv \zeta(\omega_i)$. In general, $\omega_1$ and $\omega_3$ define a fundamental parallelogram characterizing the doubly-periodic nature of the Weierstrass functions. We choose a fundamental rectangle, with $\omega_1={\bf K}(\nu)$ and $\omega_3=i\,{\bf K}^\prime\equiv i\,{\bf K}(1-\nu)$. The periods are then parametrized by the elliptic parameter $\nu$ that takes values in $[0, 1]$. At the limits $\nu$=0 and $\nu=1$, the Jacobi elliptic function reduce to trigonometric and hyperbolic functions respectively.  Physically, they interpolate between kink-like solutions and sinusoidal ones. 

The Weierstrass functions are then related to the Jacobi elliptic functions as follows.
We define
\begin{eqnarray}
\sigma_i(z)=e^{-\eta_iz} \frac{\sigma(z+\omega_i)}{\sigma(\omega_i)}
\label{sigmai}
\end{eqnarray}
and Jacobi's elliptic functions can be constructed from their ratios,
\begin{eqnarray}
{\rm sn}(z)=\frac{\sigma(z)}{\sigma_3(z)}
\quad ; \quad 
{\rm cn}(z)=\frac{\sigma_1(z)}{\sigma_3(z)}
\quad ; \quad
{\rm dn}(z)=\frac{\sigma_2(z)}{\sigma_3(z)}
\label{jacobi}
\end{eqnarray}
where the Jacobi functions have elliptic parameter $\nu$ and the Weierstrass functions have periods $\omega_1={\bf K}(\nu)$ and $\omega_3=i{\bf K}^\prime(\nu)$.

Weierstrass's zeta function is defined as the logarithmic derivative of the sigma function:
\begin{eqnarray}
\zeta(z)=\frac{d}{dz}\ln(\sigma(z))=\frac{\sigma^{\prime}(z)}{\sigma(z)}
\label{zeta}
\end{eqnarray}
From (\ref{sigmaperiod}) it is clear that $\zeta$ is also quasi periodic
\begin{equation}
\zeta(z+2 \omega_i)=2\eta_i+ \zeta(z)\qquad ; \quad i=1, 2, 3 \quad .
\label{zetaperiod}
\end{equation}
Here $\eta_i$ is given by the zeta function evaluated on the periods $z=\omega_i$:
 \begin{equation}
\zeta(\omega_i)=\eta_i
\label{eta}
\end{equation}
Finally,  the Weierstrass ${\mathcal P}$ function is defined as:
\begin{equation}
{\mathcal P}(z)=-\frac{d\zeta(z)}{dz}
\label{P}
\end{equation}
${\mathcal P}$ is doubly periodic, with periods $2\omega_1$, 2$\omega_3$:
\begin{equation}
{\mathcal P}(z+2 \omega_i)= {\mathcal P}(z)\qquad ; \quad i=1, 2, 3 \quad .
\label{Pperiod}
\end{equation}
Another important property of ${\mathcal P}$ is that it satisfies the differential equation
\begin{eqnarray}
{\mathcal P}^{\prime\,2}(z)&=&4{\mathcal P}^3(z)-g_2P(z)-g_3\nonumber\\
&\equiv&4\left({\mathcal P}(z)-e_1\right)\left({\mathcal P}(z)-e_2\right)\left({\mathcal P}(z)-e_3\right)
\label{Pdiff}
\end{eqnarray}
This equation is the one (\ref{general-amp}) satisfied by the amplitude squared of the condensate, following from the nonlinear Schr\"odinger equation. 
The constants $g_2$ and $g_3$ in (\ref{Pdiff}) are known as the $invariants$ and they are parameters depending on the periods. Similarly for the $e_i$, with $e_1+e_2+e_3=0$, and $g_2=-4(e_1e_2+e_2e_3+e_1e_3)$ and $g_3=4e_1e_2e_3$. With our choice of periods, these can be related to the Jacobi elliptic parameter $\nu$ as:
\begin{eqnarray}
e_1=\frac{1}{3}(2-\nu)\quad ; \quad 
e_2=\frac{1}{3}(2\nu-1)\quad ; \quad 
e_1=-\frac{1}{3}(1+\nu)\quad .
\end{eqnarray}

Just as the trigonometric and hyperbolic functions satisfy addition and product formulae, so too do the elliptic functions.
Indeed, these elliptic identities generate all others as special cases.
Here we list some of the important addition formulas for $\sigma$, $\zeta$ and ${\mathcal P}$ that we have used throughout the paper. References \cite{akhiezer,lawden} are particularly good concerning these identities.
\begin{eqnarray}
\frac{\sigma(u+v)\sigma(u-v)}{\sigma^2(u)\sigma^2(v)}=-{\mathcal P}(u)+{\mathcal P}(v)
\label{addition1}
\end{eqnarray}
\begin{eqnarray}
\frac{\sigma(u+v)\sigma(u-v)\sigma(2x)}{\sigma(u+x)\sigma(u-x)\sigma(v+x)\sigma(v-x)}=\zeta(u+x)-\zeta(u-x)-\zeta(v+x)+\zeta(v-x)
\label{addition2}
\end{eqnarray}
\begin{eqnarray}
\left[\zeta(u+v)-\zeta(u)-\zeta(v)\right]^2={\mathcal P}(u+v)+{\mathcal P}(u)+{\mathcal P}(v)
\label{addition3}
\end{eqnarray}
\begin{eqnarray}
\zeta(u+v)-\zeta(u-v)-2\zeta(v)=\frac{{\mathcal P}^{\prime}(v)}{{\mathcal P}(v)-{\mathcal P}(u)}
\label{addition4}
\end{eqnarray}


\end{document}